\documentclass[letterpaper, 10 pt, conference]{ieeeconf}  % Comment this line out if you need a4paper

\IEEEoverridecommandlockouts                              % This command is only needed if 
\usepackage[T1]{fontenc}
\usepackage{cite}
\usepackage{amsmath,amssymb,amsfonts}
\usepackage{algorithmic}
\usepackage{graphicx}
\usepackage{subcaption}
\usepackage{textcomp}
\usepackage{xcolor}
\usepackage[colorlinks=true, linkcolor=blue, citecolor=blue, urlcolor=blue]{hyperref}
\usepackage{placeins}
\newtheorem{lemma}{Lemma}%[section] 
\newtheorem{remark}{Remark}%[section]
\newtheorem{problem}{Problem}%[section]
\newtheorem{definition}{Definition}%[section]
\newtheorem{proposition}{Proposition}%[section]
\newtheorem{assumption}{Assumption}%[section]
\DeclareMathOperator*{\argmin}{arg\,min} 
 
\usepackage{multirow}
\usepackage{float}
\usepackage{placeins}
\usepackage{adjustbox}
\usepackage{epstopdf}
\usepackage{multicol}
\usepackage{paracol}
\usepackage{soul}
\usepackage{multirow}
\usepackage{makecell}
\usepackage{threeparttable}
\soulregister\cite7
\soulregister\eqref7

\usepackage{xurl}
\usepackage[labelformat=parens]{subcaption}
\title{\LARGE \bf
Continuous-Time Control Synthesis for Multiple Quadrotors under Signal Temporal Logic Specifications 
}
\author{Yating Yuan*, Yu Liu*%
\thanks{Yating Yuan is with the Department of Applied Mathematics, University of Waterloo, 200 University Avenue, Waterloo, Ontario, Canada N2L3G1. Yu Liu is with the College of Urban Transportation and Logistics, Shenzhen Technology University, 3002 Lantian Road, Shenzhen, Guangdong, China, 518118. *The corresponding authors. (email: {\tt\small yating.yuan@uwaterloo.ca; liuyu@sztu.edu.cn}).}%
}

\begin{document}

\maketitle
\thispagestyle{empty}
\pagestyle{empty}

%%%%%%%%%%%%%%%%%%%%%%%%%%%%%%%%%%%%%%%%%%%%%%%%%%%%%%%%%%%%%%%%%%%%%%%%%%%%%%%%
\begin{abstract}
Continuous-time control of multiple quadrotors in constrained environments under signal temporal logic (STL) specifications is critical due to their nonlinear dynamics, safety constraints, and the requirement to ensure continuous-time satisfaction of the specifications. To ensure such control, a two-stage framework is proposed to address this challenge. First, based on geometric control, a Lyapunov-based analysis of the rotational tracking dynamics is performed to facilitate multidimensional gain design. In addition, tracking-error bounds for subsequent STL robustness analysis are derived. Second, using the tracking-error bounds, a mixed-integer convex programming (MICP)-based planning framework with a backward-recursive scheme is developed. The framework is used to generate reference trajectories that satisfy multi-agent STL tasks while meeting the trajectory requirements imposed by geometric control. Numerical simulations demonstrate that, compared with uniform gains, the optimized multidimensional gains yield less conservative time-varying bounds, mitigate oscillations, and improve transient performance, while the proposed framework ensures the satisfaction of multi-agent STL tasks in constrained environments with provable tracking guarantees.

\end{abstract}


%%%%%%%%%%%%%%%%%%%%%%%%%%%%%%%%%%%%%%%%%%%%%%%%%%%%%%%%%%%%%%%%%%%%%%%%%%%%%%%%
\section{INTRODUCTION}
\label{sec:introduction}

As drone technology advances, quadrotors are increasingly used to execute complex tasks in confined environments, such as narrow passages and regions with tight terminal constraints~\cite{best2024multi}. To formally specify such tasks, signal temporal logic (STL) provides a language over continuous signals with explicit temporal semantics. However, guaranteeing continuous-time STL satisfaction for nonlinear systems remains an open challenge~\cite{belta2019formal}, especially for multi-agent systems that require both coordination and formal guarantees~\cite{sun2022multi}.

Various approaches have been proposed to ensure STL satisfaction in multi-agent systems. Among these, model predictive control (MPC) has been widely adopted due to its efficiency and ability to handle complex constraints and uncertainty~\cite{pant2018fly, zou2019event, zhou2022distributed, buyukkoccak2021distributed, charitidou2024distributed}. In~\cite{pant2018fly}, an MPC-based controller is employed for multi-agent systems subject to continuous-time STL specifications. A stricter specification is defined by tightening temporal intervals and predicates~\cite{fainekos2006robustness}. Smooth robustness is also adopted to improve efficiency in multi-agent tasks~\cite{pant2017smooth}, although this relaxation may compromise soundness~\cite{gilpin2020smooth}. Another line of work applies distributed MPC (DMPC) to cooperative control in multi-agent systems by solving local problems with neighbor communication, thereby offering a balance between scalability and coordination~\cite{zou2019event, zhou2022distributed, buyukkoccak2021distributed, charitidou2024distributed}. While effective in improving efficiency in multi-agent systems, these methods generally lack formal guarantees of continuous-time STL satisfaction. In particular, existing efforts address this challenge by assuming piecewise-constant inputs and exploiting the Lipschitz continuity of robustness functions to extend discrete-time results to continuous time~\cite{charitidou2024distributed}. However, the guarantees obtained under these assumptions remain conditional and conservative, and the iterative nature of DMPC introduces additional computational cost.

Alternatively, closed-loop strategies such as control barrier functions (CBFs) and prescribed performance control (PPC) directly enforce continuous-time STL satisfaction through real-time state-feedback control. Decentralized time-varying CBFs with switching logic are proposed in~\cite{lindemann2019decentralized}. However, guarantees of fixed-time constraint satisfaction are not provided, since the time required for constraint satisfaction is not uniformly bounded over the set of initial conditions. Fixed-time CBFs (TFCBFs) with least-violation optimization are introduced in~\cite{sharifi2021fixed} to handle task conflicts. Higher-order CBFs combined with feedback linearization are employed in~\cite{rao2023temporal} for temporal waypoint tracking. PPC provides guarantees for prescribed transient and steady-state performance~\cite{bu2023prescribed}, and it has also been extended to multi-agent systems under STL tasks. Formation and sequencing tasks are addressed in~\cite{lindemann2017prescribed}, leader-follower coordination is considered in~\cite{chen2022funnel}, and a contract-based decomposition for distributed implementation is introduced in~\cite{liu2025controller}. While these methods enable feedback-based task enforcement in continuous time, they remain limited to specific specifications, require careful parameter tuning, and may suffer from input discontinuities under switching.

Additionally, nonlinear feedback controllers with bounded tracking-error guarantees have been integrated with STL planning methods to ensure continuous-time satisfaction in~\cite{fan2020fast} and~\cite{pant2021co}. However, such approaches tend to be conservative in narrow environments due to the use of uniform tracking-error bounds. Moreover, for quadrotor applications, conventional controllers formulated in Euclidean coordinates, such as decoupled position-attitude controllers based on Euler angles, may suffer from coordinate singularities and inconsistent position-attitude coordination. In this context, geometric control on $\mathrm{SE}(3)$~\cite{gu2025robust, lee2010geometric, lee2012robust, invernizzi2017geometric, gamagedara2019geometric, 11395993} avoids coordinate singularities, thereby enabling agile maneuvers and direct handling of quadrotor dynamics.

\begin{figure}
    \centering
    \includegraphics[width=\linewidth]{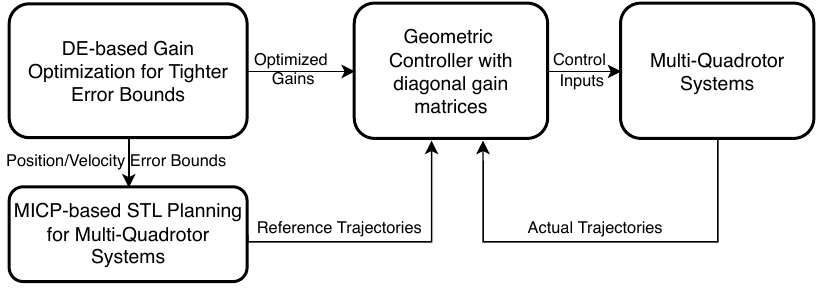}
    \caption{Configuration of the proposed framework with continuous-time STL robustness guarantees.}
    \label{fig:STL-GC_framework}
\end{figure}

% \textbf{Contributions:} 
In this work, we propose a control-synthesis framework based on geometric control, as illustrated in Fig.~\ref{fig:STL-GC_framework}. Differential evolution (DE) is employed to tune the diagonal gain matrices, yielding time-varying bounds on the position and velocity errors. These bounds are then incorporated into the MICP-based STL planner to generate reference trajectories such that the actual trajectories satisfy the STL specifications. The main contributions of this paper are summarized as follows:
\textit{1)} A geometric controller with multidimensional gains is developed by generalizing the scalar gains in~\cite{11395993} to diagonal positive-definite gain matrices. Based on a Lyapunov analysis of the tracking errors and a differential-evolution-based procedure for gain selection, the proposed method yields less conservative error bounds that decay exponentially. Numerical simulations further demonstrate faster convergence, reduced oscillations, and improved transient performance. \textit{2)} The continuous-time STL framework with Bézier curves proposed in~\cite{yu2024continuous} is extended to multi-agent systems by incorporating a backward-recursive structure~\cite{leung2023backpropagation} into the MICP encoding, thereby improving computational efficiency. Both the reference and tracked trajectories satisfy the same multi-agent continuous-time STL specifications, while the reference trajectories additionally satisfy the requirements imposed by the geometric controller.

The rest of the paper is organized as follows. Section~\ref{sec:Preliminaries} introduces the preliminaries on quadrotor dynamics, geometric control, Bézier curves, and STL. Section~\ref{sec:PS} formulates the control-synthesis problem under STL specifications. Section~\ref{sec:expdecayerror} analyzes the error bounds under the initial conditions and presents a differential-evolution-based optimization for gain selection. Section~\ref{sec:refTrajConstruction} presents the MICP-based optimization problem for generating reference trajectories for multiple quadrotors. Simulation results and illustrative examples are provided in Section~\ref{sec:exp}, and Section~\ref{sec:conclusion} concludes the paper.

\section{Preliminaries}
\label{sec:Preliminaries}

\textbf{Notation:} Let $\mathbb{Y} \subset \mathbb{R}^3$ denote the workspace. Let $\mathbf{1}_d \in \mathbb{R}^d$ denote the all-ones vector, and let $\mathrm{I}d, 0{d \times d} \in \mathbb{R}^{d \times d}$ denote the identity and zero matrices, respectively.
Let
$
\mathbb{R}_{++}^d:=\left\{x \in \mathbb{R}^d \mid x_i>0, \forall i=1, \ldots, d\right\}
$
be the set of strictly positive $d$-dimensional real vectors.
For brevity, $\mathbb{R}_{++}:=\mathbb{R}_{++}^1$ denotes the set of strictly positive real numbers, and $\mathbb{S}_{++}^n$ denotes the set of $n \times n$ symmetric positive-definite matrices. For any $a \in \mathbb{R}^d$, $|a|$ denotes the element-wise absolute value, and $\|a\|$ is the Euclidean norm. The maximum and minimum eigenvalues of a matrix $A$ are denoted by $\bar{\lambda}(A)$ and $\underline{\lambda}(A)$, and $\operatorname{tr}(A)$ denotes its trace. Let $T>0$ define the time horizon, and $\mathbb{T}=[0, T]$. For a signal $y: \mathbb{T} \rightarrow \mathbb{Y}$, its portion to an interval $[t_i, t_j]$ is defined as $\left(y,\left[t_i, t_j\right]\right):=\left\{y(t) \mid t \in\left[t_i, t_j\right]\right\}$. For brevity, the suffix of the signal $y$ starting at time $t$ is denoted by $(y, t):=(y,[t, T])$. Define a convex polytope $\operatorname{Poly}(H, b):=\left\{x \in \mathbb{R}^d \mid H x \leq b\right\}$, with $H \in \mathbb{R}^{r \times d}$ and $b \in \mathbb{R}^r$. Let $H_i$ and $b_i$ denote the $i$-th row of $H$ and entry of $b$, respectively. The number of rows in $H$ is denoted $\operatorname{Row}(H)$.

\subsection{Quadrotor Dynamics Model}
\label{subsec.uav_model}

\begin{table}[!t]
\caption{Physical variables in quadrotor dynamics.}
\label{table: Common physical variables}
\renewcommand{\arraystretch}{1.2}
\begin{center}
\begin{tabular}{|l||l|}
\hline Symbol & Physical meaning \\
\hline \hline
$\mathcal{I}$ &  Inertial frame $\mathcal{I} =\left\{e_1, e_2, e_3\right\}$\\
$\mathcal{B}(t)$ & Body frame  $\mathcal{B}(t)=\left\{b_1(t), b_2(t), b_3(t)\right\}$\\
$m \in \mathbb{R}$ & Total mass  \\
$g \in \mathbb{R}$ & Gravitational acceleration $g=9.81$ m/s$^2$\\
% \multirow{2}{*}{$J \in \mathbb{R}^{3 \times 3}$}  &  the positive-definite inertia matrix, namely, \\
% & $J=\operatorname{diag}([J_1, J_2, J_3]),J_1, J_2, J_3>0.$\\
$J \in \mathbb{S}_{++}^{3}$  &  Positive-definite inertia matrix \\
$p(t) \in \mathbb{R}^3$  
& Position of the center of mass \\
$v(t) \in \mathbb{R}^3$  & Velocity of the center of mass \\
$R(t) \in \mathrm{SO}(3)$ & Rotation matrix with $b_i(t)=R(t) e_i, i \in \{1,2,3\}$ \\
$\omega(t) \in \mathbb{R}^3$ & Angular velocity in the body-fixed frame \\
$f_i(t) \in \mathbb{R}$ & Thrust of the $i$-th propeller along $b_3(t)$\\
$f(t) \in \mathbb{R}$ &  Total thrust $f(t) = \sum^4_{i=1} f_i(t)$\\
$\tau(t) \in \mathbb{R}^3$ & Torque\\
\hline
\end{tabular}
\end{center}
\end{table}

\begin{figure}[!htb]
    \centering
\includegraphics[width=0.73\linewidth]{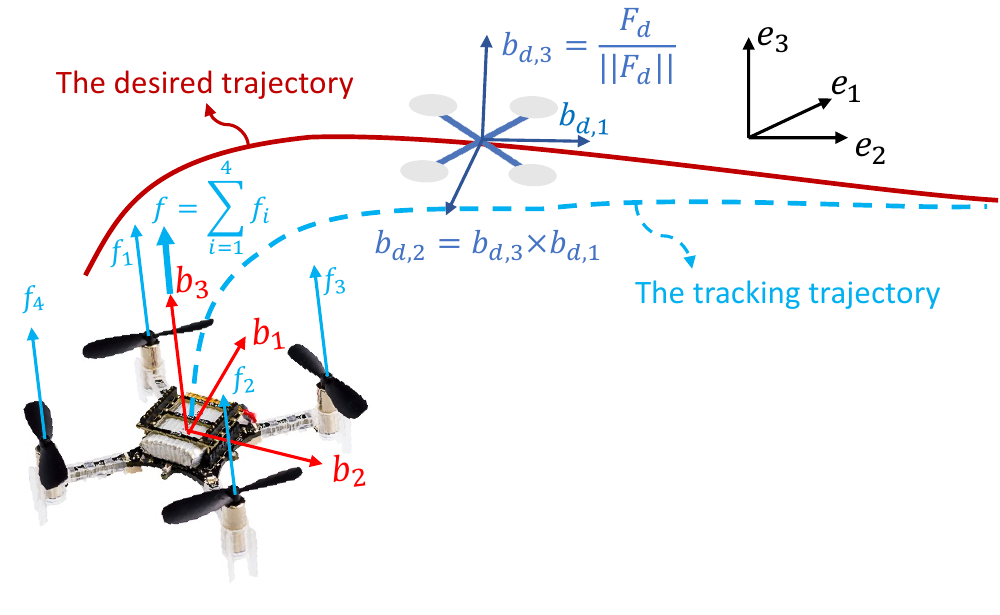}
    \caption{The inertial frame $\mathcal{I}$, the body frame $\mathcal{B}$, and desired body axes $b_{d,1}(t)$, $b_{d,2}(t)$, $b_{d,3}(t)$ representing the desired orientation $R_d(t)$.}
    \label{fig:frames}
\end{figure}

Table \ref{table: Common physical variables} lists the physical variables of the quadrotor dynamics model. As depicted in Fig.~\ref{fig:frames}, let $\mathcal{I}={e_1,e_2,e_3}$ denote an inertial frame, where $e_1$ and $e_2$ span the horizontal plane and $e_3$ points upward, opposite to the direction of gravity. Let $\mathcal{B}(t)=\{b_1(t),b_2(t),b_3(t)\}$ be the body-fixed frame attached to the quadrotor center of mass, where $b_1(t)$, $b_2(t)$, and $b_3(t)$ denote the forward, lateral, and thrust directions, respectively.

The configuration space of the quadrotor is the special Euclidean group $\mathrm{SE}(3)$, parameterized by a position $p \in \mathbb{R}^3$ and a rotation matrix $R \in \mathrm{SO}(3)$. The Lie algebra $\mathfrak{so}(3)$ of $\mathrm{SO}(3)$ is the space of skew-symmetric matrices. The hat map $\wedge: \mathbb{R}^3 \rightarrow \mathfrak{so}(3)$ is defined by $x \mapsto \hat{x}$, where $\hat{x} y=x \times y$ for all $y \in \mathbb{R}^3$. The vee map $(\cdot)^{\vee}: \mathfrak{s} \mathfrak{o}(3) \rightarrow \mathbb{R}^3$ denotes the inverse of $\wedge$ \cite[Chapter~3]{lynch2017modern}. 

The quadrotor dynamics are given by \cite{lee2010geometric}:
\begin{align}
\dot{p}(t) & =v(t), \label{eq.dp}\\
\dot{v}(t) & =-g e_3+m^{-1} f(t) R(t) e_3, \label{eq.dv}\\
\dot{R}(t) & =R(t) \hat{\omega}(t), \label{eq.dR}\\
\dot{\omega}(t) & =J^{-1}(-\omega(t) \times J \omega(t)+\tau(t)), \label{eq.dw}
\end{align}

\subsection{Geometric Control}
\label{subsec:GC}
Let $y_d:\mathbb{T} \rightarrow \mathbb{Y}$ be a desired $\mathcal{C}^4$ position trajectory\footnote{A curve is $\mathcal{C}^k$ if it is $k$-times continuously differentiable. Note the fourth derivative of the reference trajectory is required to compute the derivative of angle velocity.}. As presented in Fig.~\ref{fig:frames}, the desired rotation matrix $R_d(t) = [b_{d,1}(t), b_{d, 2}(t), b_{d,3}(t)] \in \mathrm{SO}(3)$ where $b_{d, 1}(t)$, $b_{d, 2}(t)$ and $b_{d, 3}(t)$ are chosen as follows~\cite{lee2010geometric}. The corresponding angular velocity is $\omega_d(t) = (R_d^{\top}(t) \dot{R}_d(t))^{\vee} \in \mathbb{R}^3$. Moreover, an assumption is imposed on the desired acceleration for the subsequent error-bound analysis.

\begin{assumption}[{\cite{11395993}}]
\label{Assump:acc_d}
 Assume that there exists a positive vector $b_{a} \in \mathbb{R}_{>0}^3$ such that the second-order derivative of $y_d$ satisfies the following element-wise absolute value inequality:
\begin{equation}
\left|mg e_3+m\ddot{y}_d(t)\right| \leq b_{a}, \forall t \in \mathbb{T}
\label{eq.|ge_3+ay|<B1}
\end{equation}
\end{assumption}

Let $y_{\text{tr}}:\mathbb{T} \rightarrow \mathbb{R}^3$ be a tracking trajectory with rotation matrix $R(t) \in \mathrm{SO}(3)$ and angular velocity $\omega(t) \in \mathbb{R}^3$. The positive gain matrices $K_p$, $K_v$, $K_R$, and $K_\omega \in \mathbb{S}_{++}^{3}$ are defined as
$K_{*}=\operatorname{diag}(k_{*_1}, k_{*_2}, k_{*_3})$, for $* \in \{p, v, R, \omega\}$. Define the position error, velocity error, weighted rotational error, and angular velocity error by $e_p(t)$, $e_v(t), e_{K, R}(t)$, and $e_\omega(t) \in \mathbb{R}^3$, respectively, and define the attitude error function by $\Psi_K(t) \in \mathbb{R}$ \cite{invernizzi2017geometric}:
\begin{align}
e_p(t) & :=y_{\text{tr}}(t)-y_d(t), \label{eq.e_p}\\
e_v(t) & :=\dot{y}_{\text{tr}}(t)-\dot{y}_d(t), \label{eq.e_v}\\
e_{K,R}(t) & :=\frac{1}{2}\left(K_RR^{\top}_d(t)R(t) -R^{\top}(t) R_d(t)K_R\right)^{\vee}, \label{eq.e_{K,R}} \\
e_\omega(t) & :=\omega(t)-R^{\top}(t) R_d(t)\omega_d(t), \label{eq.e_omega}\\
\Psi_K(t)  &:= \frac{1}{2}\operatorname{tr}\left(K_R(\mathrm{I}_3-R_d^{\top}(t) R(t))\right).
\label{eq.Psi(t)}
\end{align}

This assumption bounds the reference trajectory’s acceleration and facilitates subsequent analysis of the error bounds.
\begin{proposition}[{\cite[Prop.~1]{invernizzi2017geometric}}]
\label{prop:Inequality_Psi_K}
Supposing $K_R$ has distinct positive diagonal entries, define
$$
\begin{aligned}
h_1&=\min \left\{k_{R_1}+k_{R_2}, k_{R_1}+k_{R_3}, k_{R_2}+k_{R_3}\right\}, \\
h_2&=\max \left\{\left(k_{R_1}-k_{R_2}\right)^2,\left(k_{R_1}-k_{R_3}\right)^2,\left(k_{R_2}-k_{R_3}\right)^2\right\}, \\
h_3&=\max \left\{k_{R_1}+k_{R_2}, k_{R_1}+k_{R_3}, k_{R_2}+k_{R_3}\right\}.
\end{aligned}
$$
If $\Psi_K(t) < \psi < h_1$ for some positive $\psi$, then 
\begin{align}
g_1\,\|e_{K,R}(t)\|^2 \leq \Psi_K(t) \leq g_2\,\|e_{K,R}(t)\|^2,
\label{eq.g_1&_2}
\end{align}
where $g_1=\frac{h_1}{h_2+h^2_3}$ and $g_2=\frac{h_3}{h_1\left(h_1-\psi\right)}$.  
\end{proposition}

\subsection{Bézier Curves}
In this work, the reference trajectory $y_d(t)$ is parameterized by an $N$-segment Bézier spline with a uniform time interval $\Delta t = \frac{T}{N}$:
\begin{equation}
y_d(t) = B_k(t), \quad t \in [t_k, t_{k+1}],
\label{eq.y(t)}
\end{equation}
where $t_k=k \Delta t$ for $k=\{0, 1, \dots, N-1\}$. Each $B_k(t)$ is an $n$-degree Bézier curve defined as
\begin{equation}
B_k(t) = \sum_{i=0}^n b_i^n(\tau) c_{k,i}, \quad \tau = \frac{t - t_k}{\Delta t} \in [0,1],
\label{eq.def_bezier}
\end{equation}
where $c_{k,i}$ is the $i$-th control point of the $k$-th Bézier curve and $b_i^n(\tau)$ is the $i$-th Bernstein polynomial of degree $n$. Since $\sum_{i=0}^n b_i^n(\tau)=1$ for any $\tau \in[0,1]$, this implies that each $B_k(t)$ lies within the convex hull $\operatorname{conv}\left\{c_{k, 0}, \ldots, c_{k, n}\right\}=\left\{\sum_{i=0}^n \lambda_i c_{k, i} \mid \lambda_i \geq 0, \sum_{i=0}^n \lambda_i=1\right\}$.

\subsection{Signal Temporal Logic}
\label{subsec.STL}
Let $x \in \mathbb{R}^3$ and let $\mu:\mathbb{R}^3 \rightarrow \mathbb{R}$ be a predicate function. An atomic predicate is defined as $\pi := \mu(x) \geq 0$. Throughout this paper, STL formulas are considered in positive normal form (PNF)~\cite{sadraddini2015robust}. The syntax of STL is defined as:
\begin{equation}
\varphi:= \pi \mid \neg \varphi \mid  \varphi_1 \wedge \varphi_2 \mid \varphi_1 \lor \varphi_2 \mid \varphi_1 \mathcal{U}_{I} \varphi_2,
\end{equation}
where $\neg$, $\wedge$, $\lor$, and $\mathcal{U}$ are negation, conjunction, disjunction, and until operators, respectively, and $I=[a, b] \subseteq \mathbb{T}$ is a bounded time interval with $a<b$. The temporal operators \textit{Eventually} ($\Diamond$) and \textit{Always} ($\square$) can be formulated by the above operators, $\Diamond_{I} \varphi :=\text{True}\; \mathcal{U}_{I}$ and $\square_{I} \varphi:=\neg\left(\Diamond_{I} \neg \varphi\right)$. 

\begin{definition}{(STL semantics~\cite{donze2010robust})} 
\label{def. continuous semantic}
Given an STL specification $\varphi$ and a continuous-time signal $y: \mathbb{T} \rightarrow \mathbb{Y}$, the semantics of STL are defined over the suffixes $(y,t)$ as 
$$
\begin{array}{lll}
(y, t) \models \pi & \Leftrightarrow & \mu(y(t)) \geq 0; \\ 
(y, t) \models \neg \varphi & \Leftrightarrow & \neg ((y, t) \models \varphi); \\
(y, t) \models \varphi_1 \wedge \varphi_2 & \Leftrightarrow & (y, t) \models \varphi_1 \wedge(y, t) \models \varphi_2; \\
(y, t) \models \varphi_1 \vee \varphi_2 & \Leftrightarrow & (y, t) \models \varphi_1 \vee(y, t) \models \varphi_2 ;\\
(y, t) \models \Diamond_{I}  \varphi & \Leftrightarrow & \exists t^{\prime} \in \tilde{I},\left(y, t^{\prime}\right) \models \varphi; \\
(y, t) \models \square_I \varphi & \Leftrightarrow & \forall t^{\prime} \in \tilde{I},\left(y, t^{\prime}\right) \models \varphi;\\
(y, t) \models \varphi_1 \mathcal{U}_{I} \varphi_2 & \Leftrightarrow & \exists t^{\prime} \in \tilde{I} \left(y, t^{\prime}\right) \models \varphi_2 \\
& & \wedge \forall t^{\prime \prime} \in\left[t, t^{\prime}\right],\left(y, t^{\prime \prime}\right) \models \varphi_1,\\
\end{array}
$$
where $\tilde{I} = (t+I) \cap \mathbb{T}$ and $t+I = [t+a, t+b]$. The notation $(y,t) \models \varphi$ denotes that the signal suffix $(y,t)$ satisfies the STL formula $\varphi$, whereas $(y,t) \not\models \varphi$ denotes that it does not. For brevity, write $y \models \varphi$ when $(y,0) \models \varphi$.
\end{definition}

To characterize STL satisfaction of each agent's trajectory, the notion of a time-varying robust trajectory in~\cite{yuan2024signal} is extended to multi-agent systems, as formalized in Definition~\ref{def:Time-varying Robustness}.

\begin{definition}[Agent-wise time-varying robust trajectory]
\label{def:Time-varying Robustness}
Given an STL specification $\varphi$ and a function $\rho^{\ell,\varphi}:\mathbb{T}\to\mathbb{R}_{+}$ for any agent $\ell\in\{1,\dots,L\}$, the trajectory $y^\ell:\mathbb{T}\to\mathbb{Y}$ is said to be time-varying robust if $y^\ell \models \varphi$, and any trajectory $\tilde y^\ell:\mathbb{T}\to\mathbb{Y}$ satisfying
\begin{equation}
\|\tilde y^\ell(t)-y^\ell(t)\| \le \rho^{\ell,\varphi}(t), \quad \forall t\in\mathbb{T},
\end{equation}
also satisfies $\varphi$.
\end{definition}

\section{Problem Statement}
\label{sec:PS}

Given an STL specification $\varphi$, the objective is to synthesize reference trajectories and a feedback controller for multiple quadrotors such that the actual tracking trajectory $y_{\mathrm{tr}}^{\ell}$ of each quadrotor $\ell$ satisfies $\varphi$. To formalize the required robustness margin, define the lower bound on the time-varying robustness as
\begin{equation}
\gamma(t):= \Gamma(t) + \gamma_c, \quad t \in \mathbb{T}.
\label{eq.def_gamma}
\end{equation}
where $\gamma_c>0$ is a constant safety margin and $\Gamma: \mathbb{T} \rightarrow \mathbb{R}_{>0}$ is a non-increasing function. In this work, $\Gamma(t)$ characterizes a time-varying bound on the position tracking error, and $\gamma_c$ serves as a physical safety margin such as the quadrotor size.

Thus, if for each agent $\ell$ and all $t \in \mathbb{T}$, the reference trajectory $y_d^{\ell}$ satisfies $\varphi$ with robustness margin $\rho^{\ell,\varphi}(t)$ and $\rho^{\ell,\varphi}(t) \geq \gamma(t) > \Gamma(t)$, then
\begin{equation}
\left\|y_{\mathrm{tr}}^{\ell}(t)-y_d^{\ell}(t)\right\| \leq \Gamma(t)<\rho^{\ell,\varphi}(t)
\Rightarrow y_{\mathrm{tr}}^{\ell} \models \varphi.
\label{eq.y_tr_sat_stl}
\end{equation}

This overall objective can be divided into the following two sub-objectives:

\textit{Objective 1:} Optimize the multidimensional control gains for the geometric controller to obtain a non-increasing time-varying bound $\Gamma(t)$ on the tracking error.

\textit{Objective 2:} Given an STL specification $\varphi$, generate a reference trajectory $y_d^{\ell}$ in~\eqref{eq.y(t)} for each agent $\ell$, such that $y_d^{\ell} \models \varphi$ with robustness margin $\rho^{\ell,\varphi}(t) \geq \gamma(t) > \Gamma(t)$, while ensuring multi-agent safety, smoothness, and compliance with the kinematic requirements imposed by the geometric controller.

\section{Tracking Error Bounds}
\label{sec:expdecayerror}

In this section, the tracking error bounds induced by the geometric controller are analyzed in Section~\ref{subsec:ErrorBoundAnalysis}. An optimization problem is formulated in Section~\ref{subsec:opt-basedCG} to select the multidimensional control gains for $\Gamma(t)$ in \textit{Objective 1}.

With multidimensional gains, the control force $f(t)$ and torque $\tau(t)$ are defined as
\begin{flalign}
&f(t) = F^{\top}_d(t)R(t) e_3, \label{eq.f} \\
&F_d(t)=-K_p e_p(t)-K_v e_v(t)+m g e_3+m \ddot{y}_d(t), \label{eq.Fd} \\
&\tau(t)= \!-e_{K,R}(t)-K_\omega e_\omega(t)+\omega(t) \times J \omega(t) \notag \\
& \quad -J\left(\hat{\omega}(t) R^{\boldsymbol{\top}}(t) R_d(t) \omega_d(t)-R^{\top}(t) R_d(t) \dot{\omega}_d(t)\right).
\label{eq.tau}
\end{flalign}
Here, for all $t \in \mathbb{T}$, it is assumed that $\left\|F_d(t)\right\| \neq 0$ and $F_{d, 3}(t) \neq-\left\|F_d(t)\right\|$, so that the desired rotation matrix $R_d(t)$ is well defined. Similar assumptions have been adopted in related studies such as \cite{lee2010geometric} and \cite{11395993}.

\subsection{Time-varying Error Bounds Analysis}
\label{subsec:ErrorBoundAnalysis}
Tracking error bounds are derived through Lyapunov analysis of the translational and rotational tracking errors.

\subsubsection{Lyapunov Function for Translation Error}
Let $z_1^\top(t):=\begin{bmatrix} e_p^\top(t) & e_v^\top(t) \end{bmatrix}$ be the translation-error vector, and let the translation Lyapunov candidate function $\mathcal{V}_1$ be 
\begin{align}
\mathcal{V}_1(t) :=z_1^{\top}(t) M_1 z_1(t), \label{eq.V_1(t)}
\end{align}
where $\nu_1 \in (0,1)$ is a tuning constant and
\begin{align}
& M_1 =\frac{1}{2} \begin{bmatrix}
K_p & c_1\mathrm{I}_3\\
c_1\mathrm{I}_3 & m \mathrm{I}_3
\end{bmatrix} \in \mathbb{S}^6_{++}, \label{eq.M_1}\\ %
&c_1 = \nu_1 \min \left\{\sqrt{m \underline{\lambda}\left(K_p\right)},\min_{i=1,2,3} \left(\frac{4 m k_{p_i} k_{v_i}}{k_{v_i}^2+4 m k_{p_i}}\right)\right\}.\label{eq.c1}
\end{align}
Then, the time derivative of $\mathcal{V}_1(t)$ is
\begin{align}
\dot{\mathcal{V}}_1(t) =  -z_1^{\top}(t) W_1 z_1(t) - \Delta_f^{\top}(t)\left[\frac{c_1}{m}\mathrm{I}_3 \; \mathrm{I}_3\right] z_1(t), \label{eq.dotV_1(t)}
\end{align}
where 
\begin{align}
& W_1 = \frac{1}{2m}\begin{bmatrix}
2c_1 K_p & c_1 K_v \\ c_1 K_v & 2m\left(K_v-c_1\mathrm{I}_3\right)
\end{bmatrix} \in \mathbb{S}^6_{++}, \\ %\label{eq.W_1}
% & \Delta_f(t) =\left\|F_d(t)\right\| \left(\left( b_{3,d}(t) - b^{\top}_{3, d}(t) b_3(t)\right)b_3(t)\right).
&\Delta_f(t)=\left\|F_d(t)\right\|\left(\left(b_{3, d}^{\top}(t) b_3(t)\right) b_3(t)-b_{3, d}(t)\right)
\end{align}

\subsubsection{Lyapunov Function for Rotational Error}
Let $z_2^\top(t):=\begin{bmatrix} e_{K,R}^\top(t) & e_\omega^\top(t) \end{bmatrix}$ be the rotational-error vector, and let the rotational Lyapunov candidate function be
\begin{align} \mathcal{V}_2(t) := \frac{1}{2} e_\omega^{\top}(t) J e_\omega(t) + \Psi_K(t) + c_2 e_{K,R}^{\top}(t) e_\omega(t), \label{eq.V_2(t)} \end{align}
where 
\begin{equation}
\begin{aligned}
&c_2 := \nu_2 \min \Bigg\{\sqrt{2 g_m \underline{\lambda}(J)}, \; \frac{\sqrt{2} \underline{\lambda}(K_\omega)}{\operatorname{tr}(K_R)}, \\
& \qquad \qquad \qquad \qquad \min_{i =1,2,3} \frac{4 J_i k_{\omega_i}}{2 \sqrt{2} J_i \operatorname{tr}(K_R) + k_{\omega_i}^2} \Bigg\},
\label{eq.c2}
\end{aligned}
\end{equation}
with a tuning constant $\nu_2\in(0,1)$ and $g_m=\min\{g_1,g_2\}$. Here, $g_1$ and $g_2$ are as defined in Proposition~\ref{prop:Inequality_Psi_K}. From \eqref{eq.g_1&_2}, $\mathcal{V}_2(t)$ is bounded by
\begin{align}
z_2^{\top}(t) & M_{2,1} z_2(t) \leq \mathcal{V}_2(t) \leq z_2^{\top}(t) M_{2,2} z_2(t),
\label{eq.V_2_bounds}
\end{align}
where $M_{2,1}, M_{2,2} \in  \mathbb{S}_{++}^6$ are defined as
\begin{equation}
\begin{aligned}
M_{2,1} = \frac{1}{2} \begin{bmatrix} 
2g_1 \mathrm{I}_3 & c_2 \mathrm{I}_3 \\
c_2 \mathrm{I}_3 & J
\end{bmatrix}, %\label{eq.M_21}
M_{2,2} = \frac{1}{2} \begin{bmatrix}
2g_2 \mathrm{I}_3 & c_2 \mathrm{I}_3 \\
c_2 \mathrm{I}_3 & J
\end{bmatrix}.  %
\end{aligned}
\label{eq.M_21M_22} 
\end{equation}
The time derivative $\dot{\mathcal{V}}_2(t)$ satisfies:
\begin{align}
\dot{\mathcal{V}}_2(t) \leq -z_2^{\top}(t) W_2 z_2(t), \label{eq.dotV_2(t)}
\end{align}
where
\begin{align}
W_2 = \begin{bmatrix}
c_2 J^{-1} & \frac{c_2}{2} K_\omega J^{-1} \\
\frac{c_2}{2} K_\omega J^{-1} & K_\omega - \frac{c_2}{\sqrt{2}} \operatorname{tr}(K_R) \mathrm{I}_3
\end{bmatrix} \in \mathbb{S}_{++}^6. \label{eq.W_2}
\end{align}

\subsubsection{Time-varying Lyapunov Function Bounds} (See~\ref{apped:Lyapunov_Bounds} for details.)
% A brief derivation for this section is provided below, and the full details are deferred to Appendix~\ref{apped:Lyapunov_Bounds}.  

From \eqref{eq.V_2_bounds} and \eqref{eq.dotV_2(t)}, together with the Rayleigh quotient inequality~\cite[Section~4.2]{horn2012matrix}, it follows that
\begin{align}
\dot{\mathcal{V}}_2(t) \leq -\underline{\lambda}\!\left(M_{2,2}^{-\frac{1}{2}} W_2 M_{2,2}^{-\frac{1}{2}}\right)\mathcal{V}_2(t).
\end{align}
Applying the comparison principle in~\cite[Section~3.4]{khalil2002nonlinear} yields
\begin{align}
\mathcal{V}_2(t) \leq e^{-\beta t}\mathcal{V}_2(0), \quad
\beta:=\underline{\lambda}\!\left(M_{2,2}^{-\frac{1}{2}} W_2 M_{2,2}^{-\frac{1}{2}}\right). \label{eq.beta}
\end{align}
Define the bound $\mathcal{L}_2(\mathcal{V}_2(0),t):=e^{-\frac{\beta}{2}t}\sqrt{\mathcal{V}_2(0)}$, then,
\begin{align}
\sqrt{\mathcal{V}_2(t)} \leq \mathcal{L}_2(\mathcal{V}_2(0),t).
\label{eq.def_L2}
\end{align}
Moreover, using the bound on $\Psi_K(t)$ in~\eqref{eq.g_1&_2} and Rodrigues' formula~\cite[Lemma~2.3]{murray2017mathematical}, one obtains an upper bound on $\|\Delta_f(t)\|$. Consequently, \eqref{eq.dotV_1(t)} yields
\begin{align}
\dot{\mathcal{V}}_1(t) \leq - (\alpha_0 - \alpha_1\sqrt{\mathcal{V}_2(t)})\mathcal{V}_1(t) + \alpha_2\sqrt{\mathcal{V}_2(t) \mathcal{V}_1(t)},
\label{eq.dot_V1_bound}
\end{align}
where
\begin{align}
\alpha_0 &= \underline{\lambda}\left(M_1^{-\frac{1}{2}} W_1 M_1^{-\frac{1}{2}}\right),  \label{eq.a0}\\ 
\alpha_1 &=\left\|\left[K_p, K_v \right] M_1^{-\frac{1}{2}}\right\| \beta^{\prime},  \label{eq.a1} \\
\alpha_2 
&=m\left\|b_a\right\|\beta^{\prime}, \label{eq.a2} \\
\beta^{\prime} &= \left\|\left[\frac{c_1}{m} \mathrm{I}_3, \mathrm{I}_3\right] M_1^{-\frac{1}{2}}\right\| \left\|\left[\mathrm{I}_3, \mathrm{0}_{3 \times 3} \right] M_{2,1}^{-\frac{1}{2}}\right\| \sqrt{\frac{4g_2}{h_1}}. \label{eq.beta'}
\end{align}
Since $\sqrt{\mathcal{V}_2(t)} \leq \mathcal{L}_2(\mathcal{V}_2(0), t)$, applying the comparison principle in~\cite[Section~3.4]{khalil2002nonlinear} to~\eqref{eq.dot_V1_bound} together with~\eqref{eq.def_L2} yields $\sqrt{\mathcal{V}_1(t)} \leq \mathcal{L}_1(\mathcal{V}_1(0), \mathcal{V}_2(0), t)$, where
\begin{align}
\mathcal{L}_1(\mathcal{V}_1(0), \mathcal{V}_2(0), t) 
&= e^{-\frac{\alpha_0 t}{2}}  
   \left[
      e^{\frac{\alpha_1 \sqrt{\mathcal{V}_2(0)}}{\beta}} \sqrt{\mathcal{V}_1(0)} 
   \right. \notag \\
&\left. + \frac{\alpha_2 \sqrt{\mathcal{V}_2(0)}}{2} 
      \int_0^t e^{\frac{1}{2}(\alpha_0 - \beta) s} ds
   \right].
\end{align}
Both $\mathcal{L}_1\left(\mathcal{V}_1(0), \mathcal{V}_2(0), t\right)$ and $\mathcal{L}_2\left(\mathcal{V}_2(0), t\right)$ decay exponentially with time, depending on the initial values
$\mathcal{V}_1(0)$ and $\mathcal{V}_2(0)$.

\begin{proposition}[Initial conditions (See~\ref{apped:IC_Bounds})]
\label{prop.ICs}
Let $y_d:\mathbb{T}\to\mathbb{R}^3$ be a $\mathcal{C}^4$ trajectory satisfying
\eqref{eq.|ge_3+ay|<B1}. Define $\psi := \underline{\lambda}(K_R)\psi_K$ and $
\psi_K \in \left(0,\frac{h_1}{\underline{\lambda}(K_R)}\right)$, so that $\psi\in(0,h_1)$. Assume that, for some $\alpha_\psi\in[0.5,1)$ and
$\overline{\mathcal{V}}_1>0$, the initial conditions satisfy
\begin{align}
\Psi_K(0) &< \alpha_\psi\,\psi, \label{eq.Psi0<psi}\\
\frac{1}{2} e_\omega^\top(0) J e_\omega(0) &\le (1-\alpha_\psi)\psi,
\label{eq.ew0<=(1-a)psi}\\
\mathcal{V}_1(0) &\le \overline{\mathcal{V}}_1.
\label{eq.v10<=barV1}
\end{align}
Then, $\Psi_K(t)<\psi$ for all $t\in\mathbb{T}$, and hence the assumptions of Proposition~\ref{prop:Inequality_Psi_K} hold. Moreover, it follows that
\begin{equation}
\mathcal{V}_2(0)\le \overline{\mathcal{V}}_2
:=
\left(
1+c_2\sqrt{\frac{2\alpha_\psi(1-\alpha_\psi)}
{\underline{\lambda}(J)\,g_1}}
\right)\psi.
\label{eq.v20<=barV2}
\end{equation}
\end{proposition}

With~\eqref{eq.v10<=barV1} and~\eqref{eq.v20<=barV2}, $\mathcal{L}_1(\overline{\mathcal{V}}_1, \overline{\mathcal{V}}_2, t) \geq \mathcal{L}_1(\mathcal{V}_1(0), \mathcal{V}_2(0), t)$ holds. From \eqref{eq.V_1(t)}, there exist the time-varying tracking error bounds $\|e_p(t)\| \leq \mathcal{L}_p(t)$ and $\|e_v(t)\| \leq \mathcal{L}_v(t)$ where
% $\|e_{K,R}(t)\| \leq \mathcal{L}_R(t)$ and $\|e_{\omega}(t)\| \leq \mathcal{L}_\omega(t)$ 
\begin{align}
\mathcal{L}_p(t) &:=\left\|\left[\mathrm{I}_3, \mathrm{0}_{3 \times 3} \right] M_1^{-\frac{1}{2}}\right\| \mathcal{L}_1(\overline{\mathcal{V}}_1, \overline{\mathcal{V}}_2, t), \label{eq.Lpt}\\
\mathcal{L}_v(t) &:=\left\|\left[\mathrm{0}_{3 \times 3} , \mathrm{I}_3\right] M_1^{-\frac{1}{2}}\right\| \mathcal{L}_1(\overline{\mathcal{V}}_1, \overline{\mathcal{V}}_2, t).\label{eq.Lvt} 
% \mathcal{L}_f(t) &:= \left\|\left[K_p, K_v\right] M_1^{-\frac{1}{2}}\right\| \mathcal{L}_1(\overline{\mathcal{V}}_1, \overline{\mathcal{V}}_2, t).\label{eq.Lvt}
% \mathcal{L}_R(t) &:= \left\|\left[\mathrm{I}_3, \mathrm{0}_{3\times 3}\right] M_{2,1}^{-\frac{1}{2}}\right\|\mathcal{L}_2(\overline{\mathcal{V}}_2, t), \label{eq.LRt}\\
% \mathcal{L}_\omega(t) &:= \left\|\left[\mathrm{0}_{3 \times 3}, \mathrm{I}_3\right] M_{2,1}^{-\frac{1}{2}}\right\|\mathcal{L}_2(\overline{\mathcal{V}}_2, t). \label{eq.Lwt}
\end{align}

Compared with the the error bounds in~\cite{11395993}, the proposed bounds are tighter, as they are derived directly from $\mathcal{L}_1(\overline{\mathcal{V}}_1,\overline{\mathcal{V}}_2,t)$ and thus reduce conservatism.

\subsection{Optimization-based Control Gains}
\label{subsec:opt-basedCG}

The high-dimensional nonlinear quadrotor dynamics make multidimensional control-gain selection complex, and unsuitable gains may lead to unnecessarily large upper bounds. According to~\eqref{eq.Lpt} and~\eqref{eq.Lvt}, control gains can be chosen by minimizing the bound $\mathcal{L}_1(\overline{\mathcal{V}}_1,\overline{\mathcal{V}}_2,t)$.

Let $\zeta=\frac{\alpha_0-\beta}{2}$. If $\zeta \neq 0$, then $\dot{\mathcal{L}}_1\left(\overline{\mathcal{V}}_1, \overline{\mathcal{V}}_2, t\right)$ is affine in $e^{\zeta t}$, and $e^{\zeta t}$ is monotone in $t$. If $\zeta=0$, then $\dot{\mathcal{L}}_1\left(\overline{\mathcal{V}}_1, \overline{\mathcal{V}}_2, t\right)$ is affine in $t$. Therefore, the equation $\dot{\mathcal{L}}_1\left(\overline{\mathcal{V}}_1, \overline{\mathcal{V}}_2, t\right)=0$ has at most one solution. If the equation admits a solution $t_s$, then it follows by direct differentiation that $\ddot{\mathcal{L}}_1\left(\overline{\mathcal{V}}_1, \overline{\mathcal{V}}_2, t_s\right)<0$. Hence, if it exists, $t_s$ is the unique stationary point and a local maximizer of $\mathcal{L}_1\left(\overline{\mathcal{V}}_1, \overline{\mathcal{V}}_2, t\right)$. Define $t^\ast:=\max\bigl(0,\min(t_s,T)\bigr)\in\mathbb{T}$, with the convention that $t_s:=0$ when no solution exists. Then define 
\begin{align}
\mathcal{L}_{1,\max}
:=
\max_{t\in\mathbb{T}}
\mathcal{L}_1(\overline{\mathcal{V}}_1,\overline{\mathcal{V}}_2,t)
=
\mathcal{L}_1(\overline{\mathcal{V}}_1,\overline{\mathcal{V}}_2,t^\ast).
\label{eq.L1_max}
\end{align}
Furthermore, to satisfy the assumption in Proposition \ref{prop:Inequality_Psi_K} that the diagonal entries of $K_R$ are distinct, the condition $\Delta_{K_R}=\sum_{1 \leq i<j \leq 3}\Delta_{ij}=3$ is imposed, where 
% $\Delta_{ij}=1$ if $\left|K_{R_i}-K_{R_j}\right| \geq \epsilon_R$; otherwise, $\Delta_{ij}=0$.
\begin{align}
\Delta_{ij}= \begin{cases}
1, & \left|K_{R_i}-K_{R_j}\right| \geq \epsilon_R \\
0, & \text{otherwise.}
\end{cases}
\label{eq.Delta K_R}
\end{align}
In this paper,  $\epsilon_R=1$. Problem~\ref{optprobm_1} is formulated to represent \textit{Objective 1} and can be solved by differential evolution~\cite{storn1997differential}, a parallelizable global optimizer for nonconvex problems that requires no initial guess.
\begin{problem}
\label{optprobm_1}
The control gains $\left\{K_p, K_v, K_R, K_\omega\right\}$ and tuning parameters $\left\{\nu_1, \nu_2\right\}$ are determined by solving the following optimization problem:
\begin{equation}
\label{problem 1}
\begin{aligned}
\underset{\substack{
K_p, K_v, K_R, K_\omega, \\
\nu_1, \nu_2
}}{\argmin} & \qquad  \mathcal{L}_{1,\max}\\
\text {such that} \quad 
& 0 <\nu_1, \nu_2<1, \\
& \Delta_{K_R} = 3, \\
% & \mathcal{L}_p(0) \leq \mathcal{L}_{p,\text{init}} \quad \text{(If applicable)},\\
\underline{k} \leq K_{p_i}, & K_{v_i},  K_{R_i}, K_{\omega_i} \leq \bar{k}, \; i \in \{1,2,3\}.
\end{aligned}
\end{equation}
Here, $0 < \underline{k} < \bar{k}$ are constants and denote the lower and upper bounds of the control gains, respectively.
\end{problem}

\section{Reference Trajectory Construction} 
\label{sec:refTrajConstruction}
% This section formulates an optimization problem over the control points of Bézier curves to ensure that the reference trajectory satisfies the STL specifications with a robustness margin, thereby guaranteeing that the tracking trajectory also adheres to the same specifications.

Once the gains are determined by solving Problem \ref{optprobm_1}, the position error bound $\mathcal{L}_p\left(\overline{\mathcal{V}}_1, \overline{\mathcal{V}}_2, t\right)$ can be obtained. Then, according to~\eqref{eq.def_gamma}, the non-increasing function $\Gamma(t)$ defining the lower bound $\gamma(t)$ is chosen $\Gamma(t) = \tilde{L}_p(t)$, where
\begin{align}
\tilde{L}_p(t)= \begin{cases}
   \mathcal{L}_p(t^*), & \text{if } t \leq t^*, \\
\mathcal{L}_p(t), & \text{otherwise}. 
\label{eq.Gamma(t)}
\end{cases}
\end{align}
To ensure that both the Bézier-parameterized reference trajectory and the tracking trajectory satisfy the STL specifications, \textit{Objective 2} is formulated as Problem \ref{optprobm_2}.

\begin{problem}
\label{optprobm_2}
Given an STL specification $\varphi$ and $L$ quadrotors, for each quadrotor $\ell\in\{1,\dots,L\}$, as defined in \eqref{eq.y(t)}, a desired position trajectory $y_d^\ell:\mathbb{T}\rightarrow\mathbb{R}^3$ is parameterized by $N$ piecewise Bézier curves of degree $n$ in \eqref{eq.def_bezier}. For any $t \in [t_k, t_{k+1}]$ and  $k \in\{0, \dots, N-1\}$, the segment $B^{\ell}_k(t)$ has the piecewise time-varying robustness measure $\rho^{\ell,\varphi}(t) = \rho_k^{\ell,\varphi}\left(\mathcal{C}^{\ell}_k, a^{\ell}_k\right) \geq \gamma(t_k)=\Gamma(t_k) +\gamma_c > 0$. Here, $\rho_k^{\ell,\varphi}\left(\mathcal{C}^{\ell}_k, a^{\ell}_k\right)$ depends on the maximum acceleration $a^{\ell}_k \in \mathbb{R}^3$ and the control points $\mathcal{C}^{\ell}_k:=\{c^{\ell}_{k, i}\}^n_{i=0}$. Let $v^{\ell}_k \in \mathbb{R}^3$ be the maximum velocity of $B^{\ell}_k$. The MICP is formulated to maximize the robustness measure while minimizing the motion effort of the reference trajectory as follows:
\begin{align}
&\argmin_{\mathcal{C}^{\ell}_k, v^{\ell}_k, a^{\ell}_k} \; \sum^L_{l=1} \sum_{k=0}^{N-1} - \mathcal{W}\rho_k^{\ell, \varphi}(\mathcal{C}^{\ell}_k, a^{\ell}_k) +\mathcal{Q}\|v^{\ell}_k\|_1 +\mathcal{R}\|a^{\ell}_k\|_1 \notag \\
&  \quad \text {such that}  \quad 
B^{\ell}_{k} \text{ and }  B^{\ell}_{k+1} \text{ satisfies $C^4$ constraints,} \notag \\
% (B^{\ell}_k)^{(4)}(t_{k+1})  = (B^{\ell}_{k+1})^{(4)}(t_{k+1}), 
% B^{\ell}_k \; \text{and}  \; B^{\ell}_{k+1}  \; \text{are} \; \mathcal{C}^4 \text{-continuous},\\
% & |(B^{\ell}_k)^{\prime}| \leq v_{max} - \mathcal{L}_v(\overline{\mathcal{V}}_1, \overline{\mathcal{V}}_2, t)\mathbf{1}_3, \\
% & |(B^{\ell}_k)^{\prime}| \leq v_{k}, \\
%  & (B^{\ell}_k)^{\prime \prime}\text{ satisfies } \eqref{eq.|ge_3+ay|<B1}  \\ %\text{ and } \eqref{eq.ay_z>=Lf}, \\
&  \qquad \qquad  \quad \; \; B^{\ell}_k \text{ satisfies kinematic constraints,} \\
&  \qquad \qquad  \qquad \rho_k^{\ell,\varphi}\left(\mathcal{C}^{\ell}_k, a^{\ell}_k\right)  \geq \gamma(t_k), \notag
% & y_d, y_{\text{tr}}\models \varphi, 
\end{align}
where $\mathcal{W},\mathcal{Q}, \mathcal{R} \in \mathbb{R}$ are user-defined weights. All quadrotors are assumed to share the same physical configuration and Lyapunov parameters. If different models are used, it suffices to adapt $\Gamma(t)$ for each model to account for the corresponding bounds  $\gamma(t)$. 
\end{problem}

\subsection{Continuity and Kinematic Constraints}

The Bézier curves are required to satisfy the continuity and kinematic constraints, which are applied uniformly to all agents. The index $\ell$ is omitted in this subsection for clarity. 
% A key property is utilized: the $\mathfrak{q}$-th derivative of a degree-$n$ Bézier curve is a degree-$(n-\mathfrak{q})$ Bézier curve.

\subsubsection{\texorpdfstring{$\mathcal{C}^4$}{C4} continuity constraints}
\label{subsec:C4_continuity}
To ensure $\mathcal{C}^4$ continuity at the junction time $t = t_{k+1}$ between consecutive Bézier segments $B_k$ and $B_{k+1}$, five constraints are imposed on their control points, expressed as follows for $\mathfrak{q} \in \{0, 1, 2, 3, 4\}$:
\begin{equation}
\sum_{\mathfrak{p}=0}^\mathfrak{q}(-1)^\mathfrak{p}\binom{\mathfrak{q}}{\mathfrak{p}} c_{k,n-\mathfrak{p}} = \sum_{\mathfrak{p}=0}^\mathfrak{q}(-1)^{\mathfrak{q}-\mathfrak{p}}\binom{\mathfrak{q}}{\mathfrak{p}} c_{k+1, \mathfrak{p}}.
\label{eq.contn_contrs}
\end{equation}
These constraints enforce the equality between the backward finite differences of the end control points of $B_k$ and the forward finite differences of the start control points of $B_{k+1}$, both evaluated at $t=t_{k+1}$

\subsubsection{Velocity constraints}
\label{subsec:vel_con}
To ensure $|\dot{y}_{\mathrm{tr}}| \leq v_{\max}$, choose $\tilde{\mathcal{L}}_v(t)=\mathcal{L}_v(t^\ast)$ for $t\le t^\ast$ and $\tilde{\mathcal{L}}_v(t)=\mathcal{L}_v(t)$ otherwise. Let $v_k$ be the maximum velocity of the $k$-th Bézier curve and impose, for all $i\in\{0,\ldots,n-1\}$, 
\begin{align}
|c_{k,i+1}-c_{k,i}| &\le \frac{v_k \Delta t}{n}, \label{eq.|cp|<v_k} \\
0 < v_k &\le v_{\max}-\tilde{\mathcal{L}}_v(t_k)\mathbf{1}_3. \label{eq.v_k<v_max}
\end{align}
Since $\dot{B}_k(t)$ is a Bézier curve of degree $(n-1)$ and its Bernstein polynomials satisfy $\sum_{i=0}^{n-1} b_i^{n-1}(\tau)=1$ for all $\tau\in[0,1]$, it follows that $|\dot{B}_k(t)| \le \frac{n}{\Delta t}\max_i |c_{k,i+1}-c_{k,i}| \le v_k$. Therefore, by the triangle inequality, it further implies that
$
\left|\dot{y}_{\mathrm{tr}}\right| \leq\left|\dot{B}_k(t)\right|+\tilde{\mathcal{L}}_v\left(t_k\right) \mathbf{1}_3 \leq v_k+\tilde{\mathcal{L}}_v\left(t_k\right) \mathbf{1}_3 \leq v_{\max }.
$

\subsubsection{Acceleration constraints}
\label{subsec:acc_con}
Likewise, let $a_k$ denote the maximum acceleration of the $k$-th Bézier curve. Define $\Delta^2 c_{k, i}:= c_{k, i+2}-2 c_{k, i+1}+c_{k, i}$ for all $i \in\{0, \ldots, n-2\}$. To bound the segment acceleration and enforce the condition in~\eqref{eq.|ge_3+ay|<B1}, the following acceleration constraints are imposed:
\begin{align}
|\Delta^2 c_{k,i}| &\le \frac{a_k \Delta t^2}{n(n-1)}, \label{eq.|cp|<a_k} \\
\left|g e_3+\frac{n(n-1)}{\Delta t^2}(\Delta^2 c_{k,i})^\top e_3\right| &\le b_a.
\end{align}

\subsection{Encoding STL Satisfaction for Multi-Agent Systems}
\label{sec:stlspec}

To encode the STL specification $\varphi$, a binary variable $z_k^{\ell, \varphi}$ is introduced to indicate whether the Bézier segment $B_k$ satisfies the specification. Specifically, if $z_k^{\varphi}$ is true, then $\left(B_k, t\right) \vDash \varphi$ for all $t \in$ $\left[t_k, t_{k+1}\right]$.

\subsubsection{Time-varying Robustness of Bézier Curves}
\label{subsec:robust_time_BC}
The STL encoding of Bézier-curve robustness from~\cite{yuan2024signal} is adopted and extended to the multi-agent setting by introducing an agent index $\ell$. As stated in \textit{Problem 2}, each segment $B_k^\ell$ has the robustness measure $
\rho_k^\varphi(\mathcal{C}_k^\ell,a_k^\ell)=r_k^\ell-\epsilon_k^\ell \ge \gamma(t_k)$. Here, $r_k^\ell \in \mathbb{R}_{>0}$ specifies the margin ensuring that the endpoints of the curve lie strictly within the required set, while $\epsilon_k^\ell \in \mathbb{R}_{>0}$ is the margin used to constrain the intermediate control points within the same set under acceleration constraints~\cite{yuan2024signal}. 

For completeness, the encoding in~\cite{yuan2024signal} is restated using the multi-agent notation adopted in this paper. Over the finite interval $\tilde{I}=(t+I)\cap\mathbb{T}$, let $k_l = k+\lfloor a/\Delta t \rfloor$ and $k_r = \min\!\big(k+1+\lfloor b/\Delta t \rfloor,\, N-1\big)$, and assume $k_l < k_r$.
\begin{align}
&{\scalebox{0.8}{$\bullet$}} \ 
\text{Predicate } \pi:   \notag \\ 
&z_k^{\ell, \pi}  = \bigwedge_{i=1}^{\operatorname{Row}(H)} \left(\bigwedge_{j=\{0,n\}} \frac{b_i-H_ic^{\ell}_{k, j}}{\left\|H_i\right\|} \geq r^{\ell}_k \right) \wedge z_{\epsilon^{\ell}_k}
    \label{eq.predicate}. \\
&{\scalebox{0.8}{$\bullet$}} \ 
\text{Negation} \neg \pi:  \notag \\
&z_k^{\ell, \neg \pi} =  \bigvee_{i=1}^{\operatorname{Row}(H)} \left(\bigwedge_{j=\{0,n\}}  \frac{H_i c^{\ell}_{k, j} - b_i}{\left\| H_i \right\|} \geq r^{\ell}_k  \right) \wedge z_{\epsilon^{\ell}_k}. \label{eq.negation} 
\end{align}
where 
\begin{equation}
\begin{aligned}
z_{\epsilon^{\ell}_k} &:= \left( \bigwedge_{j=\{1,n\}} |c^{\ell}_{k,j} - c^{\ell}_{k, j-1}| \leq \frac{a^{\ell}_k \Delta t^2}{2n} \right)\\
    &\quad \bigwedge \left(\bigwedge^d_{j=0}\frac{8\epsilon^{\ell}_k}{3\sqrt{d} \Delta t^2}-|a^{\ell}_{k,j}| \geq 0 \right).
\label{eq.z_epsilon}
\end{aligned}
\end{equation}
\begin{align}
    {\scalebox{0.8}{$\bullet$}} \ 
    &\text{Conjunctions} \ \varphi_1 \wedge \varphi_2: z_k^{\ell,\varphi_1 \wedge \varphi_2} = z_k^{\ell,\varphi_1} \wedge z_k^{\ell,\varphi_2}. & \label{eq.conj}  \\
    {\scalebox{0.8}{$\bullet$}}  \ &\text{Disjunctions} \ \varphi_1 \vee \varphi_2: z_k^{\ell,\varphi_1 \vee \varphi_2} = z_k^{\ell,\varphi_1} \vee z_k^{\ell,\varphi_2}.  & \label{eq.disjun}  \\
    {\scalebox{0.8}{$\bullet$}}  \ &\text{Always}  \ \square_I \varphi: z_k^{\ell,\square_{I} \varphi} = \wedge_{k^\prime=k_l}^{k_r}z_{k^\prime}^{\ell,\varphi}. 
    & \label{eq.always} \\
    {\scalebox{0.8}{$\bullet$}} \ & \text{Eventually} \ \Diamond_{I} \varphi:  
    z_k^{\ell,\Diamond_{I} \varphi} = \vee_{k^\prime=k_l}^{k_r}z_{k^\prime}^{\ell,\varphi}. 
   & \label{eq.eventually}\\
    \scalebox{0.8}{$\bullet$} \ & \text{Until} \ \varphi = \varphi_1 \mathcal{U}_{I} \varphi_2: & \ \notag \\
    & \qquad \qquad  z_k^{\ell,\varphi} = \bigvee_{k^{\prime}=k_l}^{k_r} \Big(z^{\ell,\varphi_2}_{k^{\prime}}  \bigwedge \wedge_{k^{\prime\prime}=k}^{k^{\prime}-1}z^{\ell,\varphi_1}_{k^{\prime\prime}}\Big). & \label{eq.Until} 
\end{align}

\subsubsection{Encoding Multi-Agent Collision Avoidance}
\label{subsec:MA_CollisonAvoid}
To ensure safety guarantees between any two agents $i$ and $j$ (where $1 \leq i<j \leq L$), the distance between their Bézier curves $B_k^i(t)$ and $B_k^j(t)$ is required to remain at least $\epsilon_{\text{inter}}$ for all $t \in\left[t_k, t_{k+1}\right]$. For simplicity, $\mathcal{C} \mathcal{H}_k^i$ and $\epsilon_k^i$ are grouped into the tuple $A_k^i=\left(\mathcal{C H}_k^i, \epsilon_k^i\right)$.
Then, the multi-agent safety specification is defined as follows: 
\begin{equation}
z_{\text{inter}}=\bigwedge_{1 \leq i < j \leq L} \operatorname{safe}\left(A_k^i, A_k^j, \epsilon_{\text{inter}}\right),
\end{equation}
where
\begin{equation}
\begin{aligned}
& \operatorname{safe}\left(A_k^i, A_k^j, \epsilon_{\text{inter}}\right) :=
\|\operatorname{cent}(B_k^i)-\operatorname{cent}(B_k^j)\|_1  \\
& \qquad \qquad  \qquad \geq \|c_{k,0}^i-c_{k,n}^i\|_1 + \|c_{k,0}^j-c_{k,n}^j\|_1  \\
& \qquad \qquad  \qquad \quad  + 2(\epsilon_k^i+\epsilon_k^j) + \epsilon_{\text{inter}}\sqrt{d}.
\end{aligned}
\label{eq.agents_safe}
\end{equation}
and $\operatorname{cent}(B^i_{k}) = \frac{1}{n+1}\sum^n_{l=0} c^i_{k, l}$ is the geometric center of the convex hull $\mathcal{CH}^i_k$. 

\begin{lemma}
If condition~\eqref{eq.agents_safe} holds for all agent pairs $(i,j)$ 
with $1 \leq i < j \leq L$ over $t \in [t_k, t_{k+1}]$, 
then the distance between any two Bézier curves satisfies
\begin{equation}
\left\|B^i_{k}(t)-B^j_{k}(t)\right\| \geq \epsilon_{\text{inter}}. 
\end{equation}
\end{lemma}
The proof can be seen in~\ref{apped:Multi-agents}.
% The proof of the soundness of the multi-agent safety can be seen in \ref{apped:Multi-agents}.

% \textit{Proof Sketch.} 
% The span of the Bézier segment $B_k^i$, defined as $\operatorname{span}\left(B_k^i\right)=\max _{0 \leq l \leq n}\left\|c_{k, l}^i-\operatorname{cent}\left(B_k^i\right)\right\|$, quantifies that segment's spatial compactness. By the convex-hull property, it follows that $\|B^i_k(t) - \operatorname{cent}(B^i_k)\| \leq \operatorname{span}(B^i_k)$ if $t \in [t_k, t_{k+1}]$. From constraints \eqref{eq.z_epsilon} and the proof in~\cite{yuan2024signal}, it follows that 
% $
% \left\|c^i_{k,l}-\operatorname{cent}\left(B^i_k\right)\right\|
% \leq 2\epsilon^i_k + \|c^i_{k,0}-c^i_{k,n}\|.
% $
% Thus, if \eqref{eq.agents_safe} holds, using the triangular inequality implies
% $
% \|B^i_{k}(t) - B^j_{k}(t)\|_1 \geq \epsilon_{inter}\sqrt{d}
% $,
% which further implies
% $\|B^i_{k}(t) - B^j_{k}(t)\| \geq \epsilon_{inter}$.                  

\begin{remark}
The multi-agent safety encoding is more restrictive because ensuring inter-agent safety requires the full convex hull of each Bézier curve. However, under acceleration constraints and with a limited number of control points, the resulting conservatism remains manageable. Moreover, the $\ell_1$-norm replaces the $\ell_2$-norm to yield convex and piecewise-linear formulations amenable to MICP solvers.
\end{remark}

\subsubsection{Complexity Analysis}
\label{subsec:complexity}
\begin{figure}[!tbp]
\vspace{3mm}
\centering
\subfloat[Semantic expansion encoding. $\varphi(0)=\bigvee_{j=0}^3 u_j$ where $u_j=\varphi_2(j) \wedge \bigwedge_{i=0}^{j-1} \varphi_1(i),\; j=\{0, 1, 2, 3\}.
$]{%
\begin{minipage}[b]{\linewidth}\centering
\includegraphics[width=0.8\linewidth]{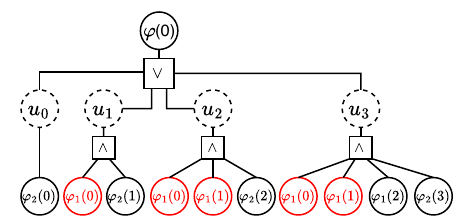}
\end{minipage}} \par\medskip
\subfloat[Backward recursion encoding. $\varphi(0)=u_{0,0}=\varphi_2(0) \vee u_{0,1}$, where the auxiliary binary $u_{i, j}$ is indexed by $i$ for $\varphi_1(i)$ and $j$ for the witness $\varphi_2(j)$, and is updated via
$
u_{i, j}=\varphi_1(i) \wedge u_{i+1, j}$, $u_{i, j-1}=u_{i, j} \vee \varphi_2(j),
$
for $i \in\{0,1,2\}$ and $j \in\{0,1,2,3\}$.]{%
\begin{minipage}[b]{\linewidth}\centering
\includegraphics[width=\linewidth]{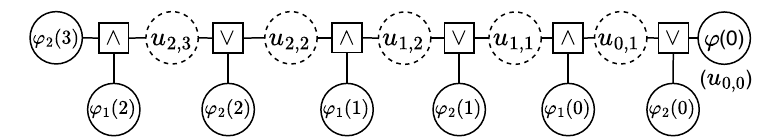}
\end{minipage}}
\caption{Two MICP encodings of $\varphi=\left(\varphi_1 \mathbf{U}_{[0,3]} \varphi_2\right)$ at time $t=0$. Dashed circles denote auxiliary binary variables. The until window uses $N_{\text{win}}=4$ discrete time steps with sampling interval $\Delta t=1$. (a) The direct semantic expansion yields an $\mathcal{O}\left(N_{\text{win}}^2\right)$-size encoding due to repeated occurrences of $\varphi_1(\cdot)$ (red circles). (b) The backward recursion shares subformulas and yields an $\mathcal{O}\left(N_{\text {win }}\right)$-size encoding.}
\label{fig:until_structures}
\end{figure}

Let $|\varphi|$ denote the number of disjunction occurrences in the STL formula $\varphi$, and let $N_{\mathrm{win}}$ denote the maximum temporal-window length measured in segment indices. For continuous-time STL specifications over piecewise Bézier trajectories with $N$ segments, binary variables are introduced only for disjunctive operators. As shown in Fig.~\ref{fig:until_structures}, the backward-recursion encoding reuses subformula encodings across parent operators, thereby avoiding the repeated encoding of overlapping subformulas. As a result, the total number of binary variables scales as $\mathcal{O}(N|\varphi|N_{\mathrm{win}})$, whereas the direct semantic expansion scales as $\mathcal{O}(N|\varphi|N_{\mathrm{win}}^2)$. For $L$ agents with $N$ Bézier segments, pairwise clearance constraints contribute $\mathcal{O}(L^2N)$ additional continuous constraints.

\section{Simulations}
\label{sec:exp}
This section evaluates the proposed framework on several benchmarks. The simulations are implemented in Python. Problem~\ref{optprobm_1} is solved using differential evolution in SciPy~\cite{2020SciPy-NMeth}, and Problem~\ref{optprobm_2} is solved as an MICP using Gurobi~\cite{gurobi}. All simulations are conducted on an 8-core Apple M1 machine with 16 GB of RAM.

\subsection{Control Gains}
The inertia matrix and mass of each quadrotor are 
$J=\operatorname{diag}(8.20,8.45,13.77)\times10^{-2}\,\mathrm{kg\cdot m^2}$ and $m=4.34\,\mathrm{kg}$, respectively, as given in~\cite{11395993}. The remaining parameters are $v_{\max}=[3,3,3]^{\top}$, $b_a=[1,1,11]^{\top}$, and $\gamma_c=0.2$.
% To select the initial-condition parameters \eqref{eq.Psi0<psi}--\eqref{eq.v10<=barV1} in Proposition~\ref{prop.ICs}, Table~\ref{tab:params_bounds} summarizes the relation between the initial values of $\psi_K$, $\alpha_\psi$, and $\overline{\mathcal{V}}_1$ and the peak position and velocity bounds, $\mathcal{L}_{\max,p}:=\max_{t\in[0,T]}\mathcal{L}_p(t)$ and $\mathcal{L}_{\max,v}:=\max_{t\in[0,T]}\mathcal{L}_v(t)$, as well as the initial-condition (IC) feasibility rate. The IC feasibility rate is defined as the percentage of feasible samples among 5000 samples drawn from the prescribed initial distributions, where $x_0,v_0\sim\operatorname{Unif}([-0.2,0.2]^3)$, $\omega_0,\mathfrak{r}_0\sim\operatorname{Unif}([-0.1,0.1]^3)$, and $R_0=\exp(\hat{\mathfrak{r}}_0)$. When tuning the parameters, one of $\psi_K$, $\alpha_\psi$, and $\overline{\mathcal{V}}_1$ is varied around the nominal setting $(\psi_K,\alpha_\psi,\overline{\mathcal{V}}_1)=(0.05,0.7,0.4)$, while the other two parameters remain fixed.

To select the initial-condition parameters \eqref{eq.Psi0<psi}--\eqref{eq.v10<=barV1} in Proposition~\ref{prop.ICs}, Table~\ref{tab:params_bounds} shows how individual variations in $\psi_K$, $\alpha_\psi$, and $\overline{\mathcal{V}}_1$ influence the peak position and velocity bounds, $\mathcal{L}_{\max,p}:=\max_{t\in[0,T]}\mathcal{L}_p(t)$ and $\mathcal{L}_{\max,v}:=\max_{t\in[0,T]}\mathcal{L}_v(t)$, as well as the initial-condition (IC) feasibility rate. The IC feasibility rate is defined as the percentage of feasible samples among 5000 samples drawn from the prescribed initial distributions, where $x_0,v_0\sim\operatorname{Unif}([-0.2,0.2]^3)$, $\omega_0,\mathfrak{r}_0\sim\operatorname{Unif}([-0.1,0.1]^3)$, and $R_0=\exp(\hat{\mathfrak{r}}_0)$. For each tuning case, one of $\psi_K$, $\alpha_\psi$, and $\overline{\mathcal{V}}_1$ is varied around the nominal setting $(\psi_K,\alpha_\psi,\overline{\mathcal{V}}_1)=(0.05,0.7,0.4)$, while the other two parameters remain fixed.

Consistent with \eqref{eq.Psi0<psi} and \eqref{eq.ew0<=(1-a)psi}, Table~\ref{tab:params_bounds} shows that increasing $\psi_K$ enlarges the overall admissible rotational initial-condition set, thereby improving IC feasibility at the cost of looser peak bounds. By contrast, for a fixed $\psi_K$, $\alpha_\psi$ mainly affects the relative strictness of the initial attitude and angular-velocity bounds. Table~\ref{tab:params_bounds} indicates that $\alpha_\psi$ has a relatively mild effect on the peak bounds, whereas the IC feasibility rate varies noticeably. As for $\overline{\mathcal{V}}_1$, increasing $\overline{\mathcal{V}}_1$ enlarges the admissible translational initial-condition set, as expected from \eqref{eq.v10<=barV1}, and this is reflected in the higher IC feasibility reported in Table~\ref{tab:params_bounds}, at the expense of larger peak bounds.

Overall, choosing $(\psi_K,\alpha_\psi,\overline{\mathcal{V}}_1)=(0.05,0.7,0.4)$ provides a balance between IC feasibility and peak-bound conservatism. The optimized control gains are given by
$$
\begin{aligned}
K_p &=\operatorname{diag}(25.2, 24.6, 25.3), && K_v=\operatorname{diag}(14.7, 14.7, 14.8), \\
K_R &=\operatorname{diag}(28.9,27.9,29.9), && K_\omega=\operatorname{diag}(2.2,1.8,2.3),
\end{aligned}
$$
and the auxiliary parameters are $\nu_1=0.75$ and $\nu_2=0.79$. Therefore, $\overline{\mathcal{V}}_2=1.46$ is obtained from \eqref{eq.v20<=barV2}, yielding the maximum position and velocity error bounds $\mathcal{L}_{\max,p}=0.61~\mathrm{m}$ and $\mathcal{L}_{\max,v}=1.46~\mathrm{m/s}$, respectively.

% Furthermore, Table~\ref{tab:params_bounds} exhibits that, compared with~\cite{11395993}, the proposed method yields tighter error bounds in most cases. This improvement can also be attributed to the use of multidimensional gain matrices, whose additional degrees of freedom allow for more effective tuning than scalar gains.
Furthermore, Table~\ref{tab:params_bounds} shows that, compared with~\cite{11395993}, the proposed method yields tighter error bounds in most cases, due to the additional tuning flexibility provided by multidimensional gain matrices.
\begin{table}[!tbh]
\centering
\begin{threeparttable}
\renewcommand{\arraystretch}{1.2}
\setlength{\tabcolsep}{2.3pt}
\caption{Effects of individual variations of $\psi_K$, $\alpha_\psi$, and $\overline{\mathcal{V}}_1$ around the nominal setting $(0.05,0.7,0.4)$ on the peak bounds $\mathcal{L}_{\max,p}$, $\mathcal{L}_{\max,v}$, and the IC feasibility rate.}
\begin{tabular}{|c|c|cc|cc|cc|}
\hline
\multirow{2}{*}{Params.} & \multirow{2}{*}{Value} & \multicolumn{2}{c|}{$\mathcal{L}_{max,p}(\operatorname{m})$} & \multicolumn{2}{c|}{$\mathcal{L}_{max, v} (\operatorname{m/s})$} & \multicolumn{2}{c|}{IC Feasibility $(\%)$} \\ \cline{3-8} 
 &  & \multicolumn{1}{c|}{Proposed} & \cite{11395993} & \multicolumn{1}{c|}{Proposed} & \cite{11395993}  & \multicolumn{1}{c|}{Proposed} &\cite{11395993} \\ \hline
 \hline
\multirow{2}{*}{$\psi_K$} & 0.005\tnote{1} & \multicolumn{1}{c|}{0.37} & $\mathbf{0.36}$  & \multicolumn{1}{c|}{0.77} &  $\mathbf{0.72}$& \multicolumn{1}{c|}{12.20} & $\mathbf{13.02}$  \\ 
\cline{2-8} 
 & 0.05 & \multicolumn{1}{c|}{$\mathbf{0.61}$ (Fig.\ref{fig:planning})} &  0.80 & \multicolumn{1}{c|}{$\mathbf{1.46}$} & 1.90 & \multicolumn{1}{c|}{$\mathbf{31.64}$} &  29.02\\ 
 \hline
 \multirow{2}{*}{$\alpha_\psi$} & 0.6  & \multicolumn{1}{c|}{$\mathbf{0.61}$} & 0.78 & \multicolumn{1}{c|}{$\mathbf{1.46}$} & 1.92 & \multicolumn{1}{c|}{$\mathbf{30.98}$} & 25.86\\ 
 \cline{2-8} 
 &  0.8 & \multicolumn{1}{c|}{$\mathbf{0.60}$} & 0.80 & \multicolumn{1}{c|}{$\mathbf{1.43}$} & 1.92 & \multicolumn{1}{c|}{$\mathbf{31.42}$} & 27.56 \\ 
\hline
\multirow{2}{*}{$\overline{\mathcal{V}}_1$} & 0.3 & \multicolumn{1}{c|}{$\mathbf{0.52}$} & 0.79 & \multicolumn{1}{c|}{$\mathbf{1.36}$} &  1.90 & \multicolumn{1}{c|}{$\mathbf{15.82}$} &$15.62$ \\ 
\cline{2-8} 
 & 0.5 & \multicolumn{1}{c|}{$\mathbf{0.69}$} & 0.80 & \multicolumn{1}{c|}{$\mathbf{1.54}$} & 1.91 & \multicolumn{1}{c|}{$\mathbf{52.28}$} &  41.42\\ \hline
\end{tabular}
\label{tab:params_bounds}
\begin{tablenotes}
\footnotesize
\item[1] $\psi_K=0.005$ follows $\bar{\Psi}$ in~\cite{11395993}, as both characterize the allowable rotational initial error.
\end{tablenotes}
\end{threeparttable}
\end{table}

The following subsection presents four scenarios to validate the proposed method. Since multi-agent safety is already encoded in $z_{\text{inter}}$, it is omitted from the subsequent specifications. For readability, the statement that robot $r$ avoids a region $\mathbf{O}$ is denoted by $r \notin \mathbf{O}$, which is logically equivalent to the negation $\neg(r \in \mathbf{O})$.

\subsection{Case 0: Runtime Analysis}
\label{subsec:case0_runtime}
\begin{figure}[!thbp]
  \centering
  % ---- table block ----
  \begin{minipage}{\linewidth}
    \centering
    \captionof{table}{Runtime (seconds) comparison in terms of the number of robots.}
    \label{tab:runtimes}
    \setlength{\tabcolsep}{6pt}
    \begin{tabular}{|c|c|c|}
      \hline
      \#Robots & The proposed & MICP \cite{yuan2024signal} \\ \hline \hline
      2 & $\mathbf{0.16}$ & 0.91 \\ \hline
      3 & $\mathbf{1.17}$ & 5.56 \\ \hline
      4 & $\mathbf{8.79}$ & 27.20 \\ \hline
      5 & $\mathbf{17.89}$ & 73.80 \\ \hline
      6 & $\mathbf{54.42}$ & 171.98 \\ \hline
    \end{tabular}
  \end{minipage}
  \vspace{1pt}
  \begin{minipage}{\linewidth}
    \centering
    \includegraphics[width=0.6\linewidth]{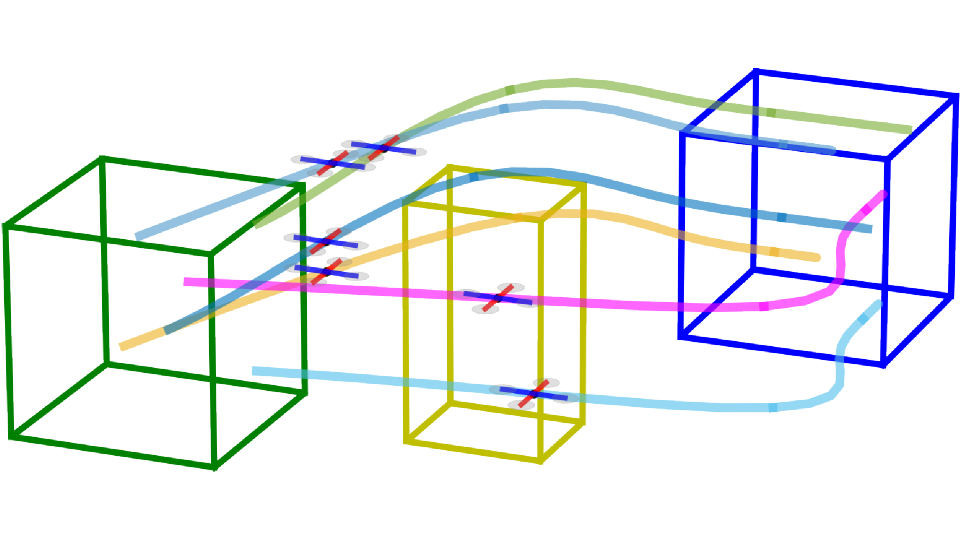}
    \caption{Case~0 environment with six quadrotors ($L=6$). }
    \label{fig:case0_sixquads}
  \end{minipage}
\end{figure}

To evaluate how runtime scales with the number of agents, we consider the simple scenario shown in Fig.~\ref{fig:case0_sixquads}. Because this scenario is simple and can accommodate different numbers of quadrotors, it better isolates the effect of $L$ on computation time. The specification is
\begin{equation}
\varphi =\bigwedge_{\ell=1}^{L}\left(\square_{[0, T]} (r_{\ell} \notin \mathbf{Y})\right) \wedge (\Diamond_{[0, T]} r_{\ell} \in \mathbf{B}),
\end{equation}
which requires the robots to avoid the yellow obstacle $\mathbf{Y}$ and reach the blue region $\mathbf{B}$. Table~\ref{tab:runtimes} shows that runtime increases with the number of agents $L$, consistent with the $\mathcal{O}(L^2N)$ scaling of pairwise collision-avoidance constraints discussed in Section~\ref{subsec:complexity}. The runtime is compared against the MICP baseline proposed in~\cite{yuan2024signal}, which uses direct semantic expansions and Bézier-curve trajectory parameterization. To enable a fair comparison in the multi-agent setting, the same collision-avoidance encoding is also incorporated into the baseline. As the number of robots $L$ increases, the proposed method shows lower runtime than the baseline. This is due to the backward-recursive encoding of temporal operators, which avoids the quadratic blow-up associated with witness-based semantic expansions.

\subsection{Case 1: Narrow Corridors}
 In Case 1, the quadrotors are required to fly through a continuously narrowing passage away from the five yellow
 $\left\{\mathbf{Y}_i\right\}_{i=1}^5$, and eventually reach a compact blue region $\mathbf{B}$, as shown in Fig.~\ref{fig:planning}(a)--(c). The navigation constraints are getting progressively tighter. Given $T = 17~\mathrm{s}$, the specification is 
\begin{equation}
\varphi =\bigwedge_{\ell=1}^2\Big(\left(\wedge_{i=1}^5 \square_{[0, T]} (r_{\ell} \notin \mathbf{Y}_i)\right) \wedge (\Diamond_{[0, T]} r_{\ell} \in \mathbf{B})\Big).
\end{equation}

\subsection{Case 2: Reach and Avoid}
Case 2 requires each quadrotor $r_{\ell}$ to stay away from the yellow obstacles $\mathbf{Y}$ and remain in the green region $\mathbf{G}$ for at least 2 seconds. Finally, all quadrotors need to reach the diagonally opposite blue region, as depicted in Fig.~\ref{fig:planning}(e)--(g), resulting in an interlaced routing challenge. Given $T=20\mathrm{~s}$, the specification is 
\begin{equation}
\begin{aligned}
\varphi = \bigwedge_{\ell=1}^4 &\Big(
\left(\square_{[0,T]} (r_\ell \notin \mathbf{Y})\right)
\wedge\left(\Diamond_{[0,T]}\square_{[0,2]} r_\ell \in \mathbf{G}\right)\\
&\quad \wedge \left(\Diamond_{[0,T]} r_\ell \in \mathbf{B}_{\ell'}\right)
\Big),
\end{aligned}
\end{equation}
where the pairing
$
\left(\ell, \ell^{\prime}\right) \in\{(1,3),(2,4),(3,1),(4,2)\}
$
encodes this diagonal goal assignment.

\subsection{Case 3: Key–Door Puzzle}
\begin{figure*}[!tbp]
% \vspace{3mm}
\centering
% ---------- Row 1 ----------
\subfloat[$t=3.6~\mathrm{s}$]{%
\begin{minipage}[b]{0.22\linewidth}\centering
\includegraphics[width=0.8\linewidth]{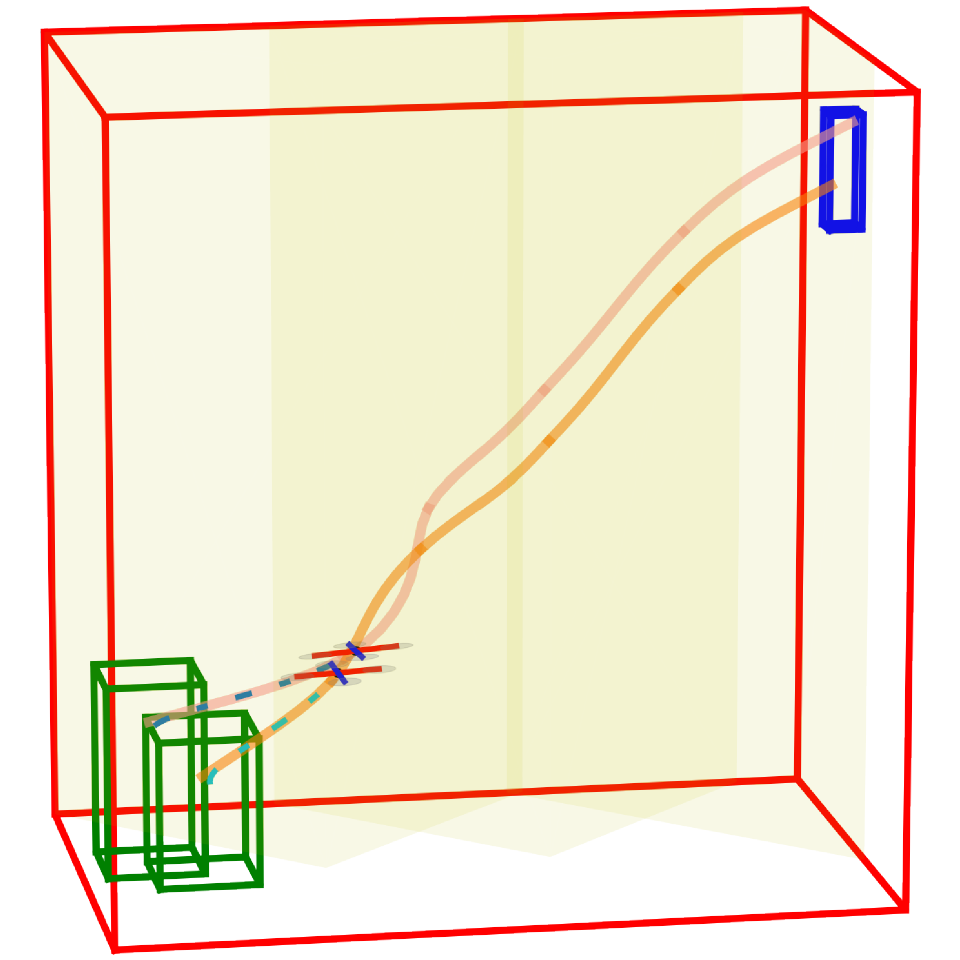}
\end{minipage}} \hfill
\subfloat[$t=7~\mathrm{s}$.]{%
\begin{minipage}[b]{0.22\linewidth}\centering
\includegraphics[width=0.8\linewidth]{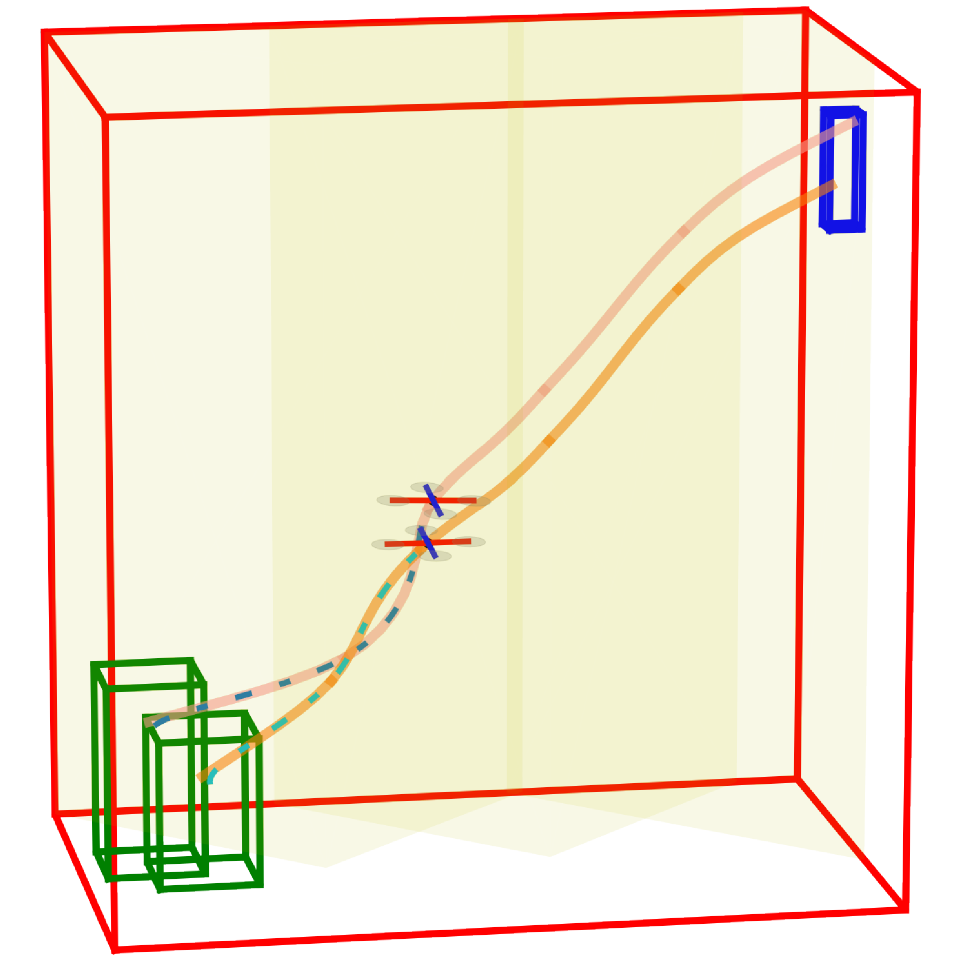}
\end{minipage}} \hfill
\subfloat[Top view.]{%
\begin{minipage}[b]{0.22\linewidth}\centering
\includegraphics[width=0.9\linewidth]{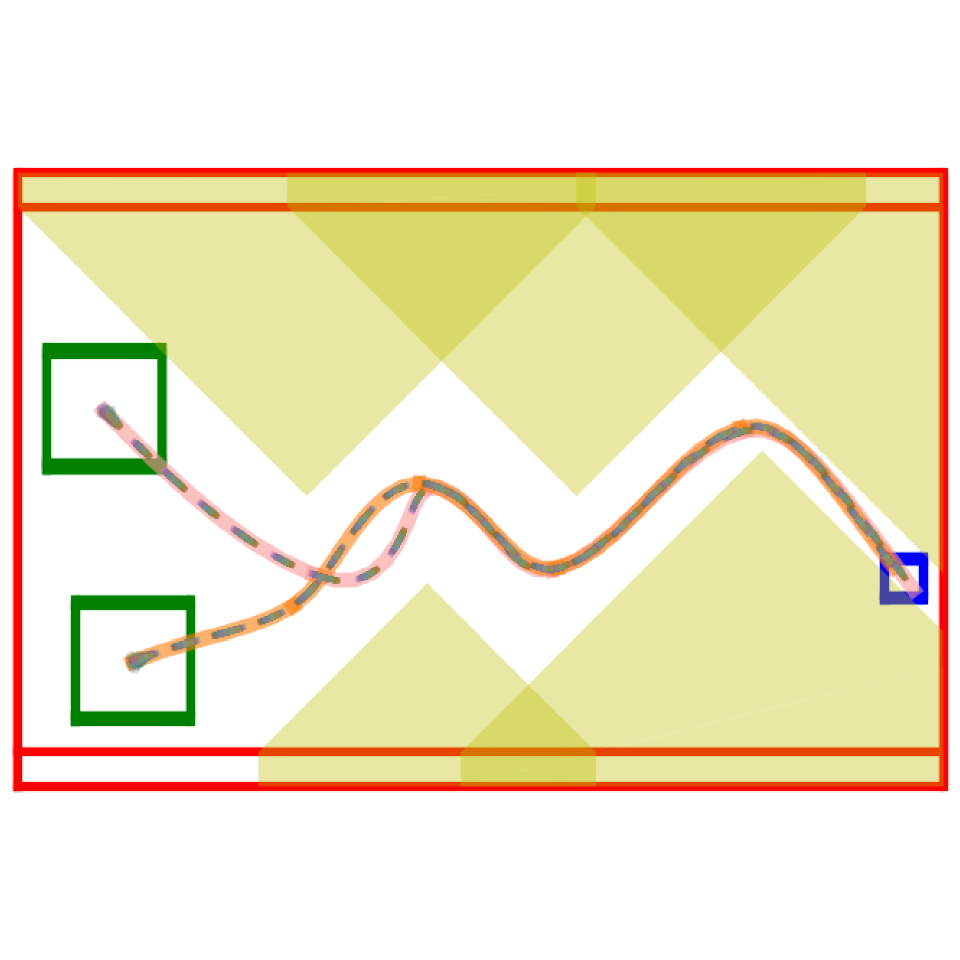}
\end{minipage}} \hfill
\subfloat[Position and velocity error norms.]{% 
\begin{minipage}[b]{0.28\linewidth}\centering
\includegraphics[width=\linewidth]{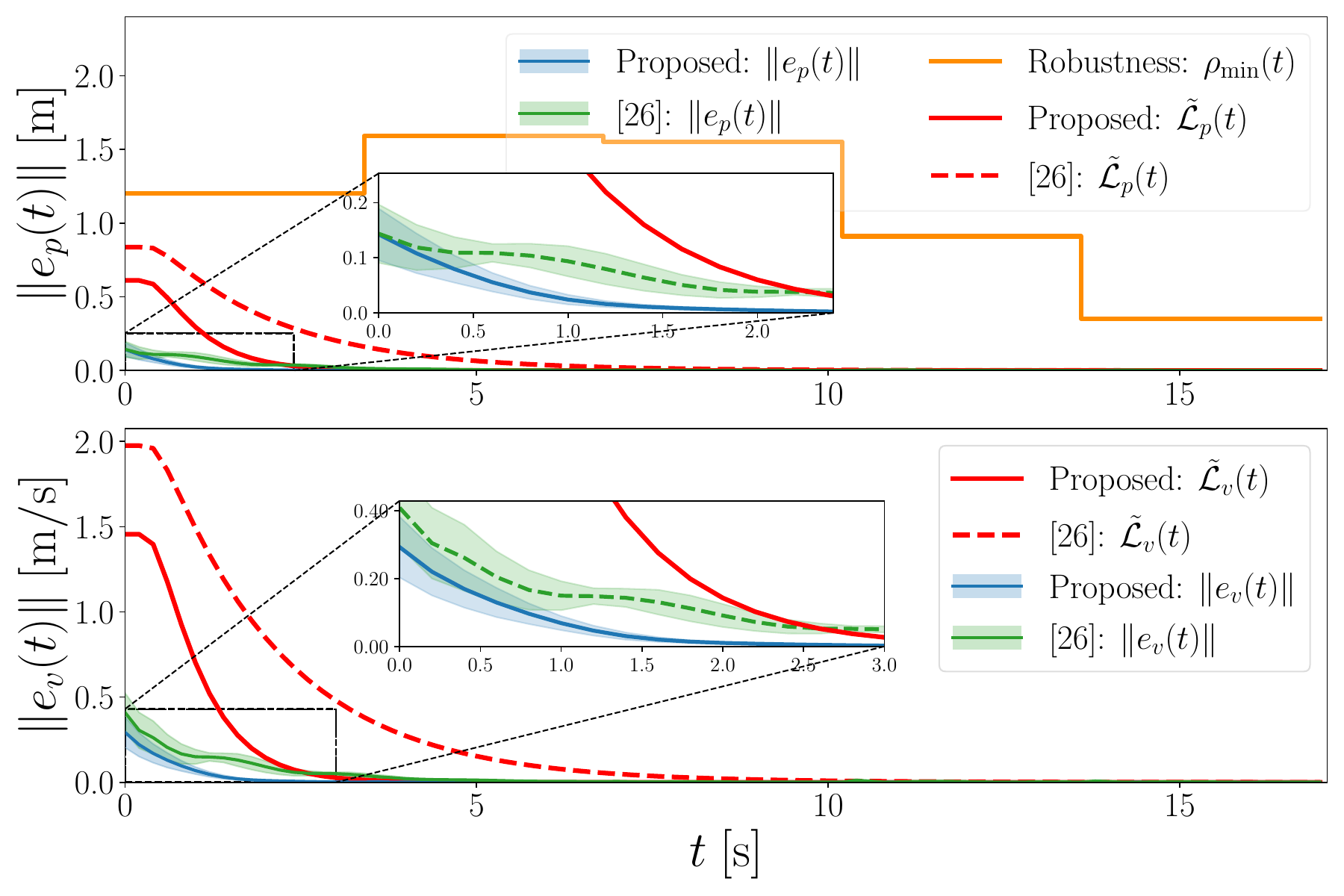}
\end{minipage}}
\\[2mm]

% ---------- Row 2  ----------
\subfloat[$t=4.2~\mathrm{s}$]{%
\begin{minipage}[b]{0.22\linewidth}\centering
\includegraphics[width=0.8\linewidth]{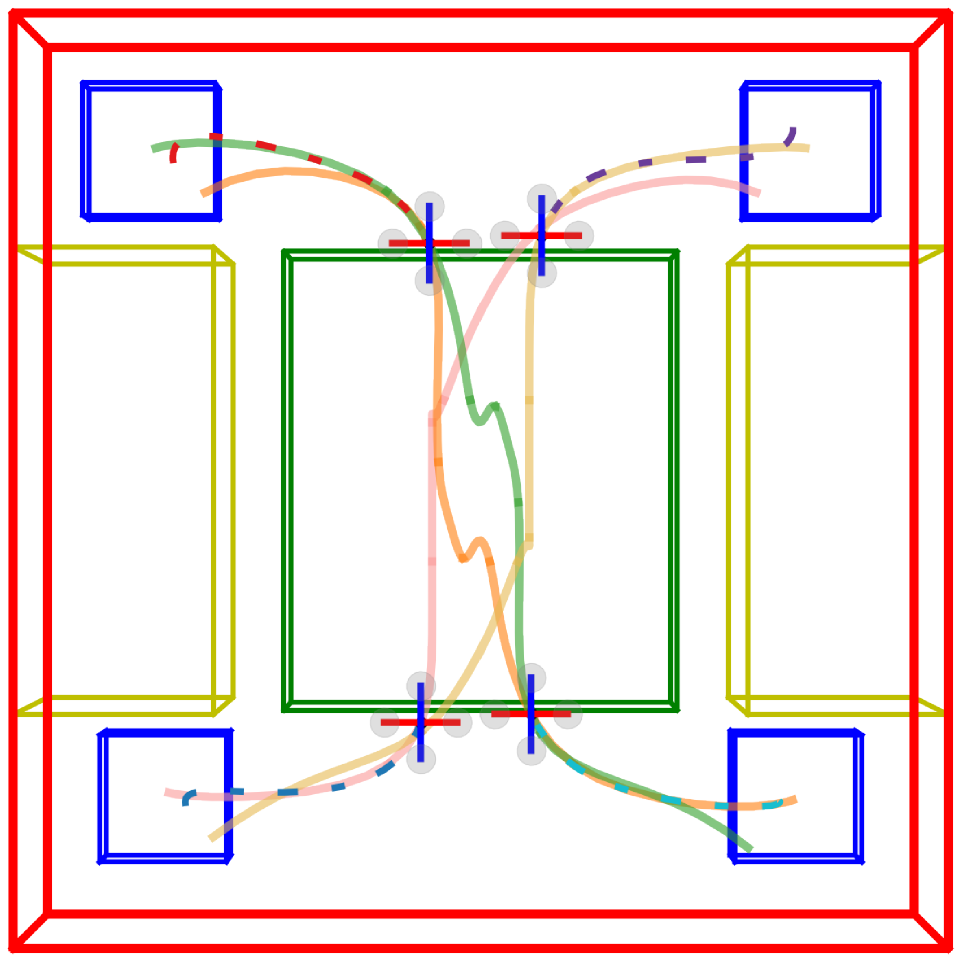}
\end{minipage}} \hfill
\subfloat[$t=9.6~\mathrm{s}$.]{%
\begin{minipage}[b]{0.22\linewidth}\centering
\includegraphics[width=0.8\linewidth]{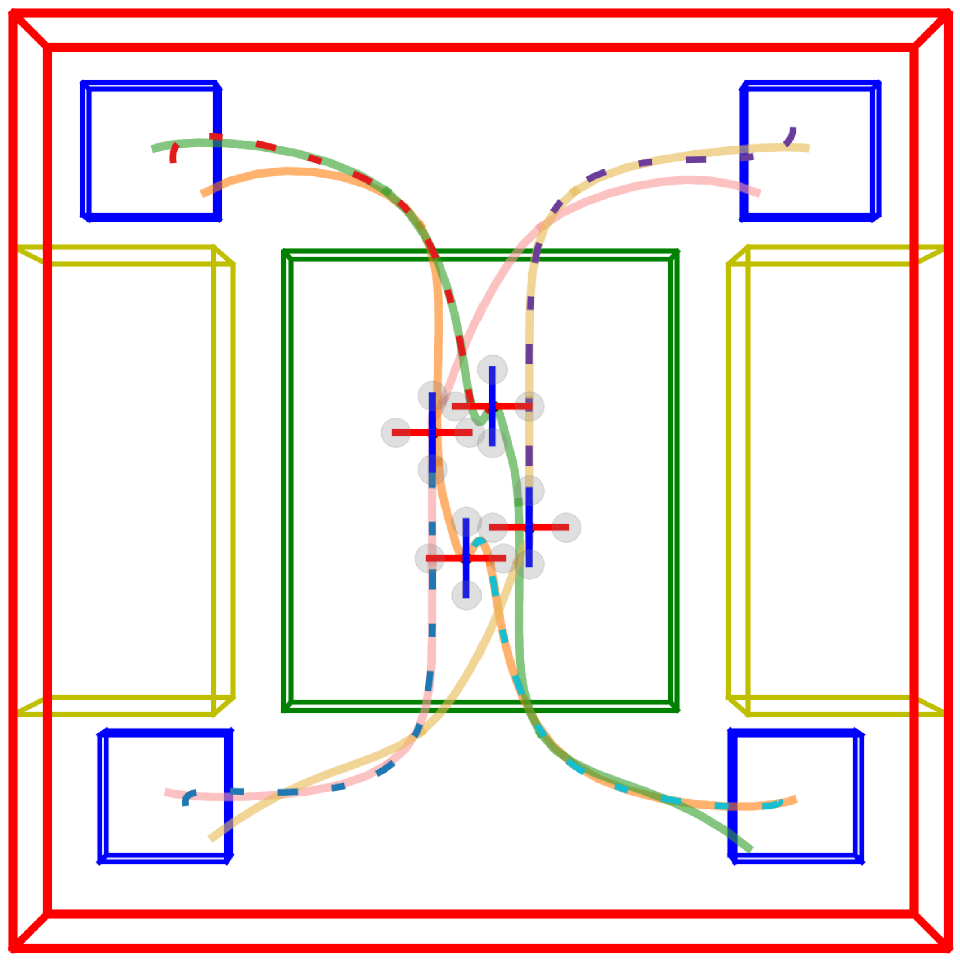}
\end{minipage}} \hfill
\subfloat[Side view.]{%
\begin{minipage}[b]{0.22\linewidth}\centering
\includegraphics[width=0.85\linewidth]{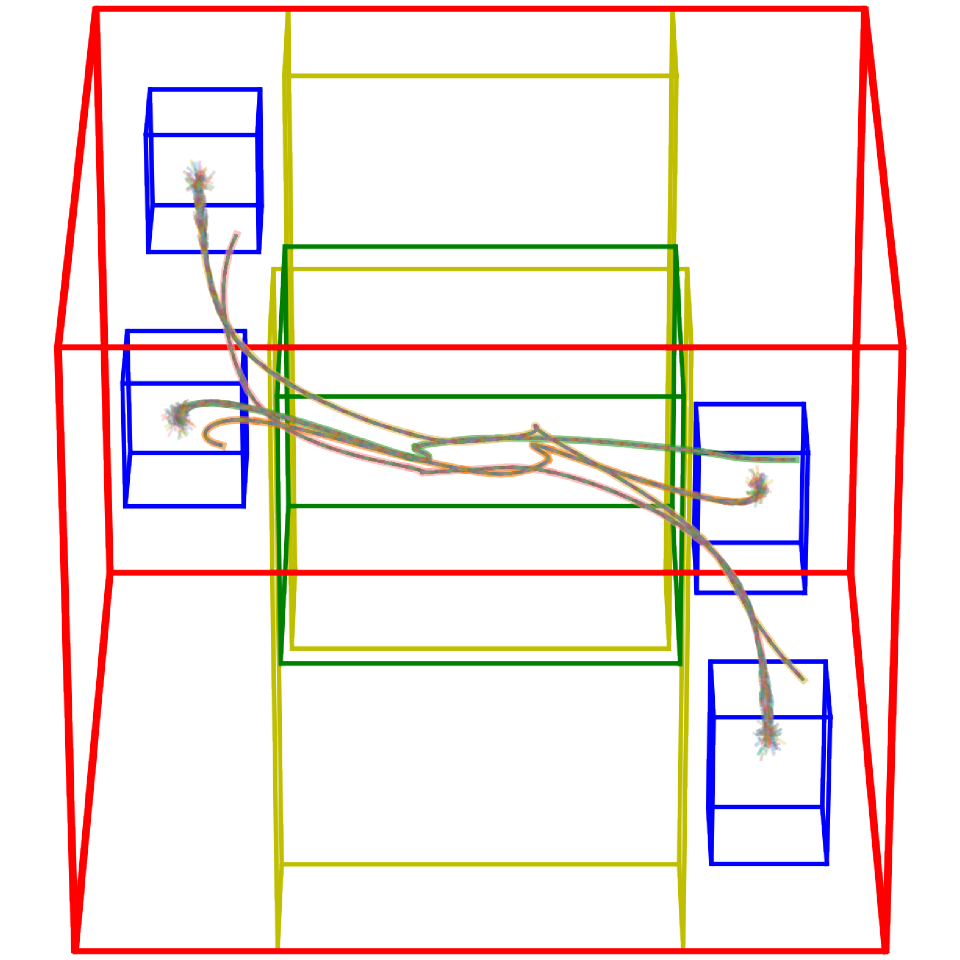}
\end{minipage}} \hfill
\subfloat[Position and velocity error norms.]{%
\begin{minipage}[b]{0.28\linewidth}\centering
\includegraphics[width=\linewidth]{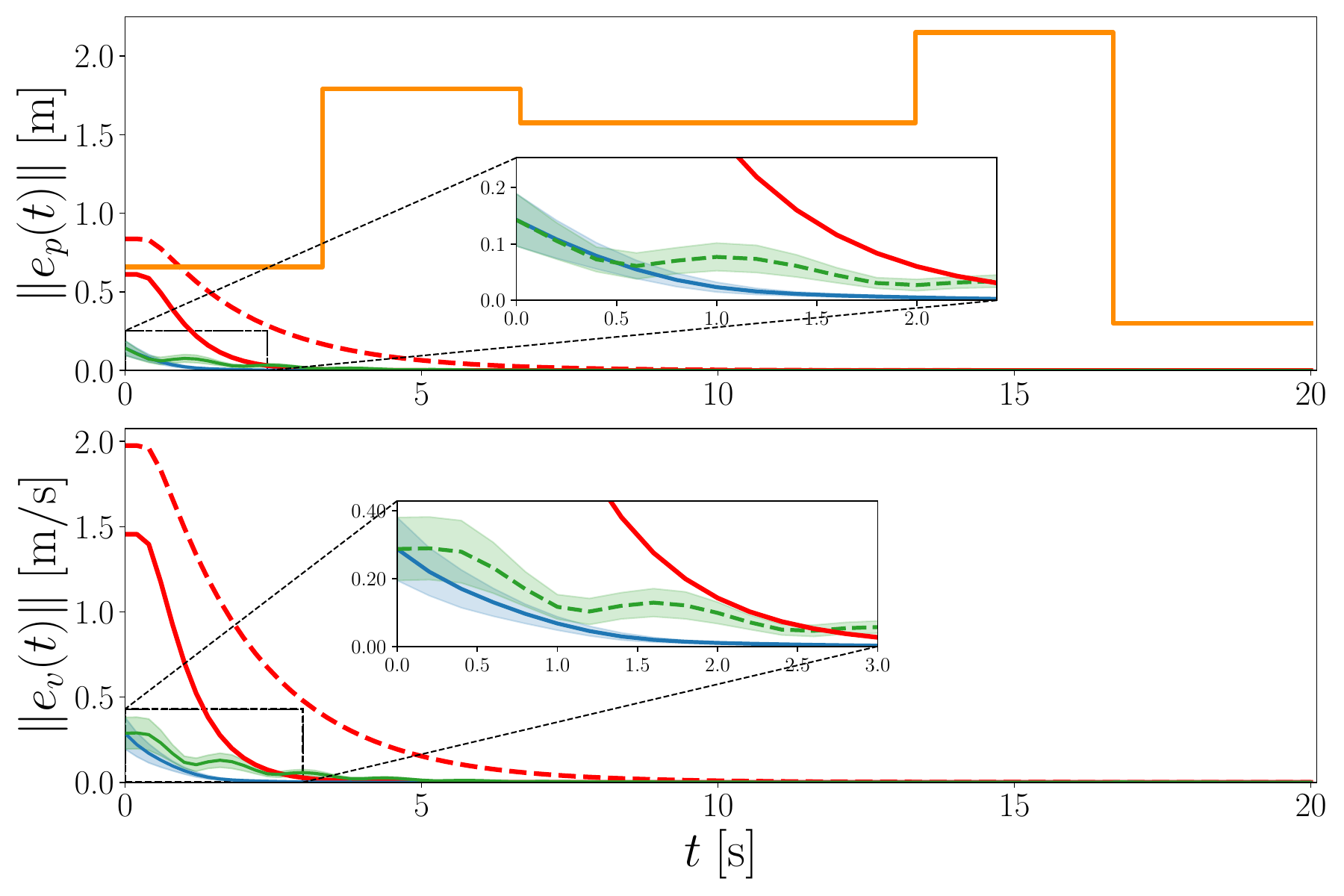}
\end{minipage}} \hfill
\\[2mm]

% ---------- Row 3 (same as Row 1) ----------

\subfloat[$t=7.6~\mathrm{s}$.]{%
\begin{minipage}[b]{0.22\linewidth}\centering
\includegraphics[width=0.8\linewidth]{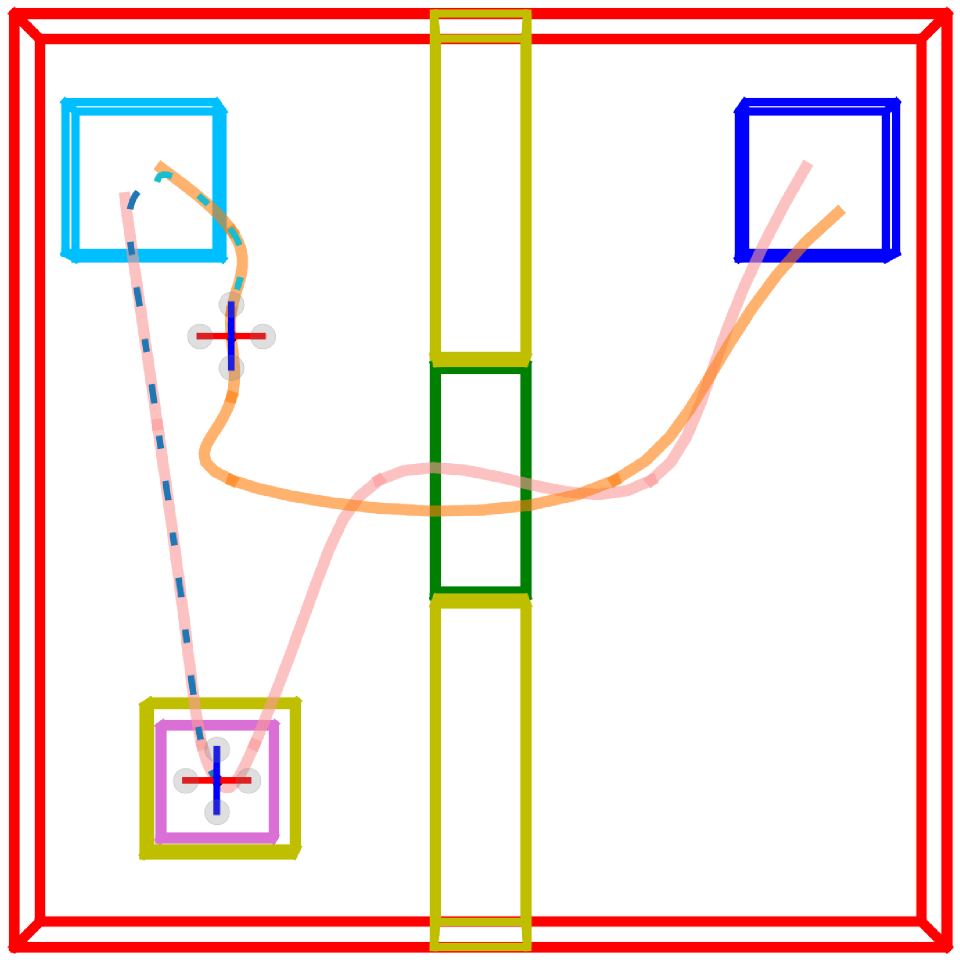}
\end{minipage}} \hfill
\subfloat[$t=14.6~\mathrm{s}$.]{%
\begin{minipage}[b]{0.22\linewidth}\centering
\includegraphics[width=0.8\linewidth]{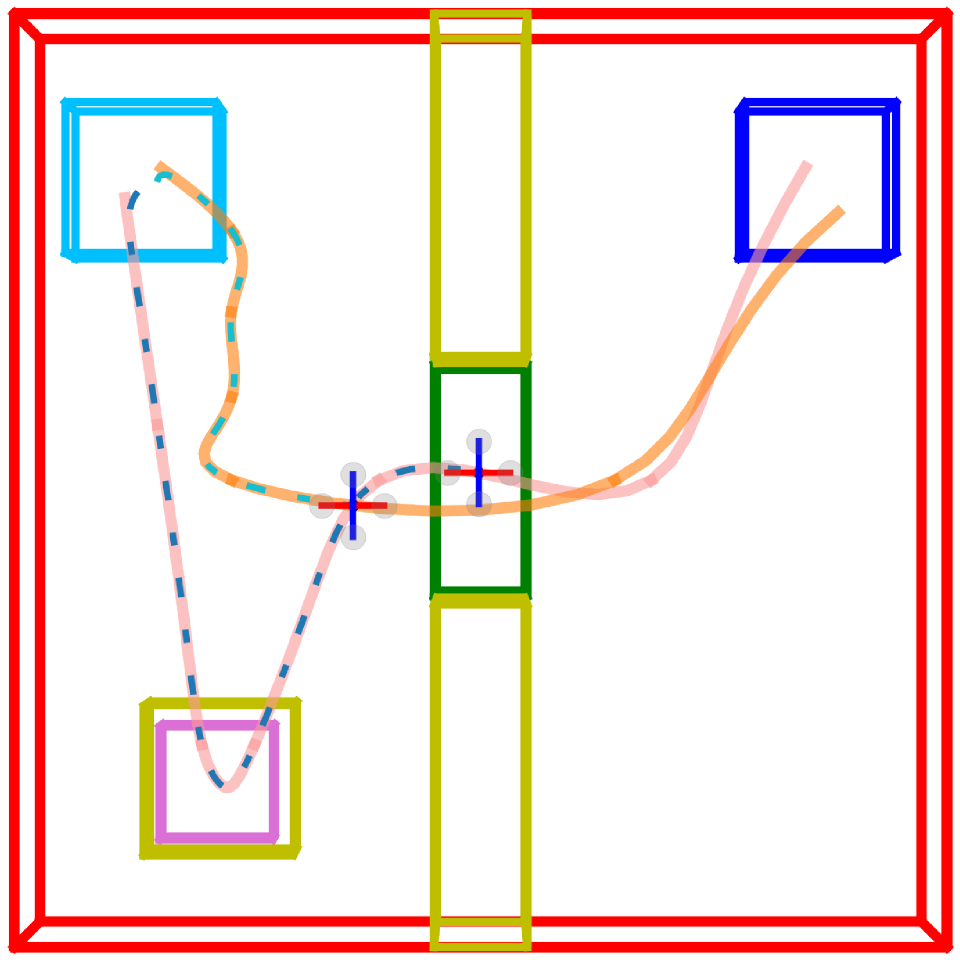} 
\end{minipage}} \hfill
\subfloat[Side view.]{%
\begin{minipage}[b]{0.22\linewidth}\centering
\includegraphics[width=0.85\linewidth]{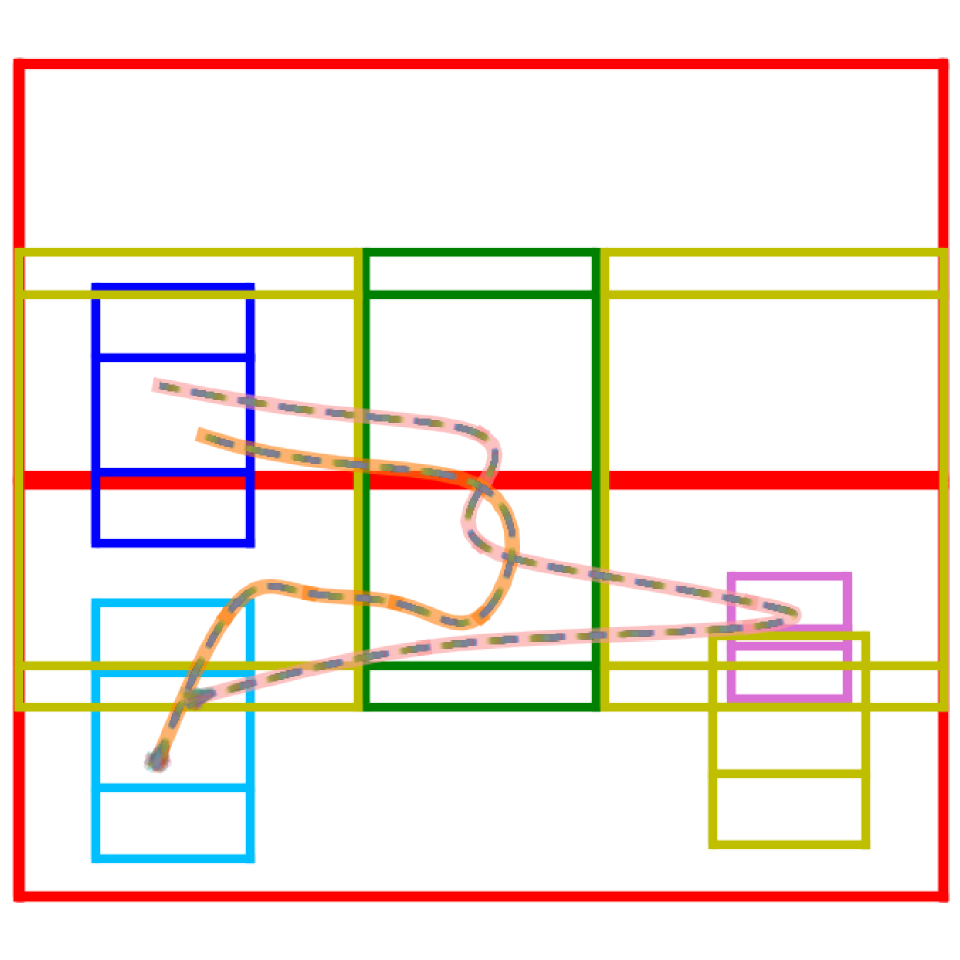} 
\end{minipage}}\hfill
\subfloat[Position and velocity error norms.]{%
\begin{minipage}[b]{0.28\linewidth}\centering
\includegraphics[width=\linewidth]{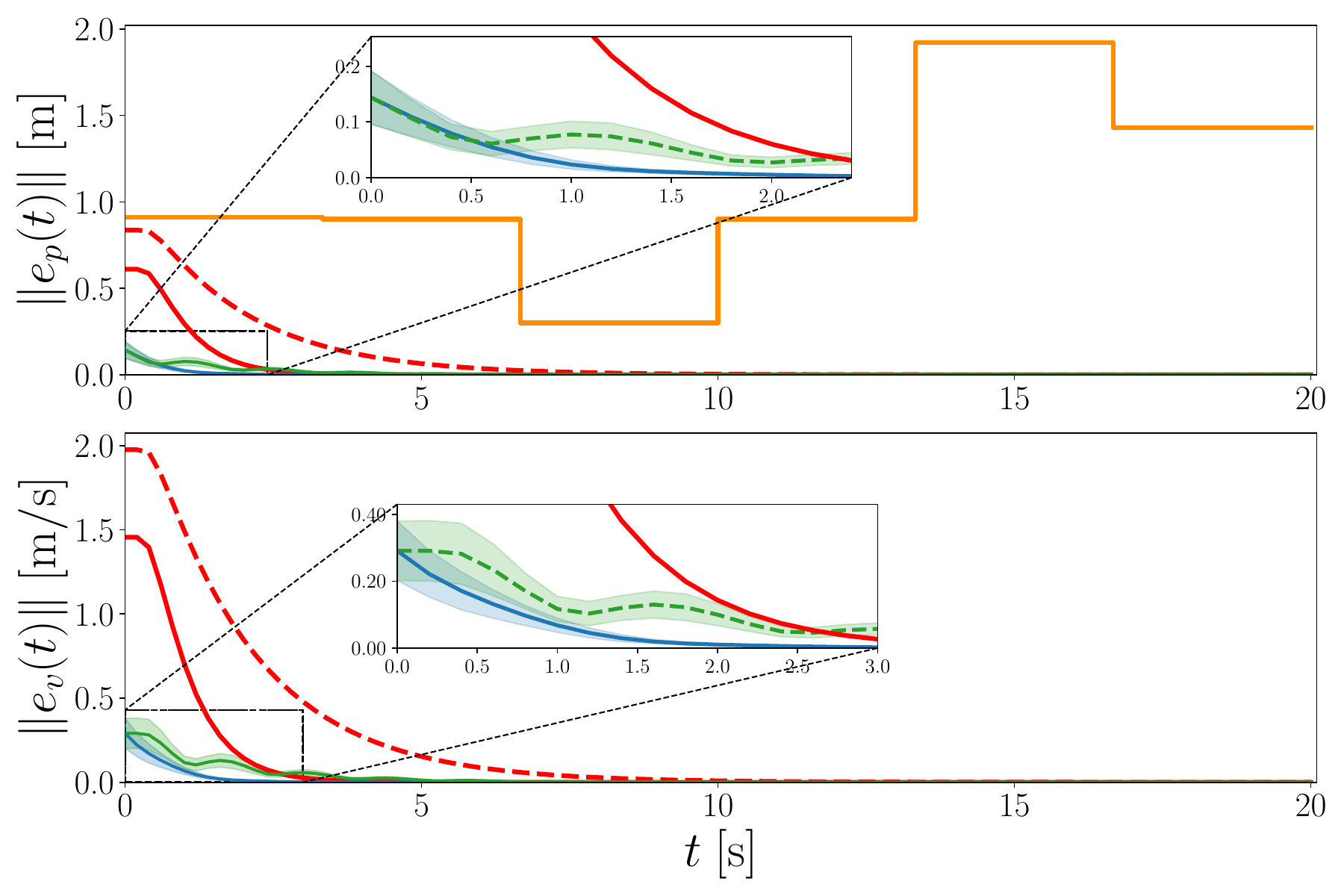}
\end{minipage}} 
% \caption{Benchmarks and results (Cases 1-3, top to bottom). In each environment plot, solid lines denote planned trajectories and dashed lines denote tracking trajectories. Columns 1-2 show two time snapshots (only the trajectory with the largest $\left\|e_p\right\|$ is plotted for clarity), illustrating inter-agent safety and STL satisfaction; for Case 1, the triangular obstacles in the side views are omitted for visibility. Column 3 provides additional views and 100 tracking trajectories. Column 4 compares the proposed method with \cite{11395993} in the position and velocity error norms $\left\|e_p(t)\right\|,\left\|e_v(t)\right\|$(mean$\pm$std, $100$ runs; lines: mean, shaded regions: std)) and their corresponding bounds $\tilde{L}_p(t), \tilde{L}_v(t)$.  The results show that the proposed method achieves less conservative bounds and faster error convergence, while ensuring $\tilde{L}_p(t)\leq \rho_{\min}(t)$, where the worst-case STL robustness margin is defined as $\rho_{\min }(t)= \min _i \rho_i(t)$.
% }
\caption{Benchmarks and results for Cases 1--3 (top to bottom). In the environment plots, solid lines denote planned trajectories and dashed lines denote tracking trajectories. Columns 1--2 show two representative time snapshots (only the trajectory with the largest $\|e_p\|$ is plotted for clarity), Column 3 shows additional views and tracking trajectories, and Column 4 compares the position and velocity tracking errors and their corresponding bounds for the proposed method and~\cite{11395993}.}
\label{fig:planning}
\end{figure*}

In Case 3, robot $r_1$ must first retrieve the key in the purple region $\mathbf{K}$ to unlock the green gate $\mathbf{G}$. Only then can both robots reach the blue target region $\mathbf{R}$ by passing through $\mathbf{G}$ while avoiding the yellow walls $\mathbf{Y}$, as shown in Fig.~\ref{fig:planning}(i)--(k). Here, with $T=20$, the specification is expressed as
\begin{equation}
\begin{aligned}
\varphi=&\left(r_2 \notin \mathbf{G} \right) \mathcal{U}_{[0, T]}\left(\left(r_1 \notin \mathbf{G}\right) \mathcal{U}_{[0, T]}\left(r_1 \in \mathbf{K} \right) \right) \\
&\bigwedge^2_{\ell=1} \left(\left(\Diamond_{[0, T]} r_{\ell} \in \mathbf{R}\right) \wedge \left(\square_{[0, T]} r_{\ell} \notin \mathbf{Y}\right) \right).
\end{aligned}
\end{equation}
This specification is  complex due to the nested structure of the two \textit{Until} operators.

In all cases mentioned above, the Bézier trajectories are parameterized by eight control points ( $n=8$ ), and the tracking time step is $d t=0.2 \mathrm{~s}$. For each case, the planned trajectories $y_d^{\ell}$ and the corresponding tracking trajectories $y_{\mathrm{tr}}^{\ell}$ are generated for all agents $\ell$. For each approach, 100 tracking trials per agent are simulated under identical initial error conditions. The experimental results are analyzed in terms of stability, safety, and robustness.

\begin{table*}[!htbp]
\centering
% \caption{Tracking performance comparison between the proposed method and~\cite{11395993} uses pooled trajectories across all agents in each scenario (100 initial conditions per agent), reporting results as mean $\pm$ std. The position-error convergence time $t_{c, p}$ is defined as the earliest time such that $\left\|e_p(t)\right\| \leq 10^{-2}~\mathrm{m}$ for all $t \geq t_{c, p}$. Post-convergence errors are time-averaged per trial as $\bar{e}_p$ over $t \geq t_{c, p}$ and $\bar{e}_v$ over $t \geq t_{c, v}$, where $t_{c, v}$ is defined analogously from $\left\|e_v(t)\right\|$.}
\caption{Tracking performance comparison between the proposed method and~\cite{11395993}, reported as mean $\pm$ std over pooled trajectories from all agents in each case. Here, $t_{c,p}$ and $t_{c,v}$ are the position and velocity error convergence times, respectively.}
\label{tab:tracking_results}
\renewcommand{\arraystretch}{2}
\setlength{\tabcolsep}{0.6pt}
\begin{tabular}{|c|c|c|c|c|c|c|c|c|c|c|}
\hline
\multirow{2}{*}{Case}
& \multicolumn{2}{c|}{\makecell{Computation time $\mathrm{(s)}$}}
& \multicolumn{2}{c|}{\makecell{Convergence time $t_{c, p}\mathrm{(s)}$}} & \multicolumn{2}{c|}{$\bar{e}_p \pm \sigma_p(\times 10^{-3} \; \operatorname{m})$} & \multicolumn{2}{c|}{\makecell{Convergence time $t_{c, v}\mathrm{(s)}$}}& \multicolumn{2}{c|}{$\bar{e}_v \pm \sigma_v\;(\times 10^{-3}\operatorname{m/s})$} \\
\cline{2-11}
& The proposed & \cite{11395993}  & The proposed & \cite{11395993}
& The proposed & \cite{11395993} 
& The proposed & \cite{11395993}
& The proposed & \cite{11395993}\\ 
\hline\hline

Narrow Corridors & $\mathbf{0.54 \pm 0.03}$ & $0.60 \pm 0.02$  & $\mathbf{1.57 \pm 0.24}$ &  $4.41 \pm 0.36$ & $\mathbf{0.60 \pm 0.07}$ & $1.73 \pm 0.10$ & $\mathbf{2.10 \pm 0.24}$ & $5.38 \pm 0.32$ & $\mathbf{0.76\pm 0.14}$ & $2.29 \pm 0.01$\\
\hline

Reach-Avoid & $\mathbf{0.48 \pm 0.02}$ & $0.52 \pm 0.03$ & $\mathbf{1.57 \pm 0.23}$ & $4.33 \pm 0.39$ & 
$\mathbf{0.38 \pm 0.04}$ & $1.02 \pm 0.08$ & $\mathbf{2.10 \pm 0.23}$ & $5.89 \pm 0.49$  & $\mathbf{0.40\pm 0.10}$ & $1.2 \pm 0.21$\\
\hline

Key-Door & $\mathbf{0.52 \pm 0.08}$  & $0.57 \pm 0.05$ &  $\mathbf{1.58 \pm 0.24}$  & $4.34 \pm 0.44$ & $\mathbf{0.37 \pm 0.02}$ & $0.97 \pm 0.01$ & $\mathbf{2.11 \pm 0.24}$ & $5.91 \pm 0.48$  & $\mathbf{0.37 \pm 0.04}$ & $1.1 \pm 0.01$\\
\hline
\end{tabular}
\end{table*}

\begin{table}[!htbp]
\setlength{\tabcolsep}{2pt}
\centering
\caption{Safety summary across cases: number of robots ($\ell$), number of segments (\#S), worst-case robustness margin 
$\rho^{\varphi}_{\min}=\min_{1\leq \ell \leq L}\min_{t_k\in\mathcal{T}}\rho^{\ell, \varphi}(t_k)$, 
and minimum inter-trajectory distances from planned ($\operatorname{dist}_{\mathrm{plan}}$) and tracking ($\operatorname{dist}_{\mathrm{track}}$) trajectories.}
\label{tab:safety_bounds}
\renewcommand{\arraystretch}{1.5}
\setlength{\tabcolsep}{3.5pt}
\begin{tabular}{|c|c|c|c|c|c|c|}
\hline
Case & $\ell$ & \#S &  Runtime$(\mathrm{s})$ &\textbf{$\rho^{\varphi}_{\min}$} & \textbf{$\operatorname{dist}_{\operatorname{plan}}$} & \textbf{$\operatorname{dist}_{\operatorname{track}}$} \\ \hline \hline
Narrow Corridors & 4 & 5 & 0.07 & 0.3 & 0.615 & 0.616\\ \hline
Reach-Avoid & 2 & 6 & 11.92&0.3 & 0.462  & 0.461 \\ \hline
Key-Door & 2 & 6 & 1.28  &0.3 & 0.773 & 0.772\\ \hline
\end{tabular}
\end{table}

% \subsection{Stability Analysis.} 

% Table~\ref{tab:tracking_results} and Fig.~\ref{fig:planning} compare the proposed multidimensional-gain geometric control method with the scalar-gain geometric control of~\cite{11395993} whereas Table~\ref{tab:safety_bounds} reports the safety evaluation.

% From the fourth column of Fig.\ref{fig:planning}, compared to~\cite{11395993}, the proposed controller yields lower mean $\left\|e_p(t)\right\|$ and $\left\|e_v(t)\right\|$ with narrower shaded bands (mean $\pm$ std over 100 initial conditions), indicating smaller dispersion across initial conditions. This trend is consistent with Table~\ref{tab:tracking_results}, which shows that across all benchmark cases the proposed method achieves shorter convergence times, smaller post-convergence mean errors, and reduced standard deviations, suggesting faster transients, tighter steady-state tracking, and fewer residual oscillations.

% Beyond tracking accuracy, we also evaluate execution efficiency. Although the proposed controller employs a vector-valued gain, its faster transient response drives the system into the low-error regime earlier, thereby shortening the time to complete each tracking run and resulting in the reduced tracking runtime reported in Table~\ref{tab:tracking_results}. Overall, the proposed approach improves both closed-loop regulation quality and tracking efficiency.

\subsection{Stability Analysis}

Table~\ref{tab:tracking_results} illustrates the comparison of the stability performance of the proposed method and that of~\cite{11395993}. The position error convergence time $t_{c,p}$ is defined as the earliest time such that $\|e_p(t)\|\le 10^{-2}\,\mathrm{m}$ for all $t\ge t_{c,p}$. Similarly, the velocity error convergence time $t_{c,v}$ is defined as the earliest time such that $\|e_v(t)\|\le 10^{-2}\,\mathrm{m/s}$ for all $t\ge t_{c,v}$. The post-convergence errors $\bar{e}_p$ and $\bar{e}_v$ are defined as the means of all finite samples of $\left\|e_p(t)\right\|$ for $t \geq t_{c, p}$ and $\left\|e_v(t)\right\|$ for $t \geq t_{c, v}$, respectively, pooled over all agents and trajectories in each case.

As shown in the fourth column of Fig.~\ref{fig:planning}, the proposed controller yields lower time-varying mean errors $\left\|e_p(t)\right\|$ and $\left\|e_v(t)\right\|$, together with narrower shaded bands, indicating smaller dispersion across trials over time. Because the post-convergence behavior is not directly evident in Fig.~\ref{fig:planning}, the convergence times and post-convergence errors in Table~\ref{tab:tracking_results} show that the proposed method achieves faster transients, tighter steady-state tracking, and reduced residual oscillations relative to the scalar-gain baseline~\cite{11395993}.

Table~\ref{tab:tracking_results} also lists the computation time. Although the proposed controller employs vector-valued gains, its faster transient response enables more efficient tracking and leads to a lower computation time for completing a reference trajectory. Overall, the proposed approach improves both tracking performance and computational efficiency.

\subsection{Safety and Robustness Analysis}
Safety is assessed by examining both the planning and tracking trajectories. A safety threshold of $\epsilon_{\text {inter}}=0.2 \mathrm{~m}$, equal to $\gamma_c$, is enforced on the minimum inter-trajectory distance. As reported in Table~\ref{tab:safety_bounds}, the minimum inter-trajectory distances are evaluated over the discrete time $\mathcal{T}=\left\{t_k=k d t \mid 0 \leq k d t \leq T\right\}$. The minimum distance between planning trajectories is defined as
\begin{equation}
\operatorname{dist}_{\mathrm{plan}} =\min_{1\le i<j\le L}\;
\min_{t_k\in\mathcal{T}} \left\|y^{i}_{d}(t_k)-y^{j}_{d}(t_k)\right\|.
\end{equation}
The tracking distance $\operatorname{dist}_{\mathrm{track}}$ is the overall minimum inter-trajectory distance over all time steps in $\mathcal{T}$, across all agents and all tracking trials (100 per agent). Table~\ref{tab:safety_bounds} shows that, in all cases, the minimum inter-trajectory distances exceed the safety threshold $\epsilon_{\text{inter}}$, thereby ensuring collision-free behavior during both planning and tracking. Moreover, the worst-case robustness across all robots is $\rho^{\varphi}_{\min}=0.3>0$, confirming that all Bézier trajectories satisfy the STL specifications. Meanwhile, Fig.~\ref{fig:planning} displays that $\rho^{\varphi}_{\min }(t) \geq \tilde{L}_p(t) = \Gamma(t)$ for all $t \in[0, T]$, implying that the tracked trajectories also satisfy the STL specifications according to~\eqref{eq.y_tr_sat_stl}.

% \section{Conclusion}
% This work proposes a continuous-time control framework for multi-quadrotor systems under STL specifications, combining geometric feedback with MICP-based Bézier planning to ensure safety, velocity limits, and STL satisfaction. The method provides provable, time-varying tracking error bounds. While effective, it depends on accurate models and initial error conditions, and incurs significant computational costs for large teams. Future work includes distributed implementations to reduce complexity, extensions to $\mathrm{SE}(3)$ curve planning for full-pose coordination, and integrating learning-based methods for improved adaptability.

\section{Conclusion}
\label{sec:conclusion}
This paper presents a continuous-time control framework for multi-quadrotor systems under STL specifications by integrating geometric tracking control with MICP-based Bézier trajectory generation. The proposed approach enables safe and dynamically feasible motion planning with formal tracking guarantees. Compared with conventional scalar-gain designs, the matrix-gain geometric controller yields tighter tracking bounds and improved tracking performance. In addition, the multi-agent extension of the continuous-time STL Bézier planning framework improves computational efficiency while ensuring specification satisfaction. However, the method still relies on accurate models and suitable initial tracking conditions, and its computational burden increases with the number of agents. Future work includes distributed implementations for improved scalability, extensions to $\mathrm{SE}(3)$ curve planning for full-pose coordination, and the integration of learning-based methods for enhanced adaptability.

\bibliographystyle{unsrt}
\bibliography{references}

@ARTICLE{11395993,
  author={Serry, Mohamed and Yuan, Yating and Chang, Haocheng and Liu, Jun},
  journal={IEEE Transactions on Automatic Control}, 
  title={Reach-Avoid Control Synthesis for a Quadrotor UAV with Formal Safety Guarantees}, 
  year={2026},
  volume={},
  number={},
  pages={1-15},
  doi={10.1109/TAC.2026.3664766}}

@article{best2024multi,
  title={Multi-robot, multi-sensor exploration of multifarious environments with full mission aerial autonomy},
  author={Best, Graeme and Garg, Rohit and Keller, John and Hollinger, Geoffrey A and Scherer, Sebastian},
  journal={The International Journal of Robotics Research},
  volume={43},
  number={4},
  pages={485--512},
  year={2024},
  publisher={SAGE Publications Sage UK: London, England}
}

@article{belta2019formal,
  title={Formal methods for control synthesis: An optimization perspective},
  author={Belta, Calin and Sadraddini, Sadra},
  journal={Annual Review of Control, Robotics, and Autonomous Systems},
  volume={2},
  number={1},
  pages={115--140},
  year={2019},
  publisher={Annual Reviews}
}

@inproceedings{fan2020fast,
author="Fan, Chuchu
and Miller, Kristina
and Mitra, Sayan",
title="Fast and Guaranteed Safe Controller Synthesis for Nonlinear Vehicle Models",
booktitle="Computer Aided Verification",
year="2020",
publisher="Springer International Publishing",
address="Cham",
pages="629--652",
isbn="978-3-030-53288-8"
}

@article{sun2022multi,
  title={Multi-agent motion planning from signal temporal logic specifications},
  author={Sun, Dawei and Chen, Jingkai and Mitra, Sayan and Fan, Chuchu},
  journal={IEEE Robotics and Automation Letters},
  volume={7},
  number={2},
  pages={3451--3458},
  year={2022},
  publisher={IEEE}
}

@inproceedings{lindemann2017prescribed,
  title={Prescribed performance control for signal temporal logic specifications},
  author={Lindemann, Lars and Verginis, Christos K and Dimarogonas, Dimos V},
  booktitle={Proc. of CDC},
  pages={2997--3002},
  year={2017},
  organization={IEEE}
}

@article{bu2023prescribed,
  title={Prescribed performance control approaches, applications and challenges: A comprehensive survey},
  author={Bu, Xiangwei},
  journal={Asian Journal of Control},
  volume={25},
  number={1},
  pages={241--261},
  year={2023},
  publisher={Wiley Online Library}
}

@inproceedings{pant2018fly,
  title={Fly-by-logic: Control of multi-drone fleets with temporal logic objectives},
  author={Pant, Yash Vardhan and Abbas, Houssam and Quaye, Rhudii A and Mangharam, Rahul},
  booktitle={2018 ACM/IEEE 9th International Conference on Cyber-Physical Systems (ICCPS)},
  pages={186--197},
  year={2018},
  organization={IEEE}
}

@inproceedings{sadraddini2015robust,
  title={Robust temporal logic model predictive control},
  author={Sadraddini, Sadra and Belta, Calin},
  booktitle={2015 53rd Annual Allerton Conference on Communication, Control, and Computing (Allerton)},
  pages={772--779},
  year={2015},
  organization={IEEE}
}

@article{yu2024continuous,
  title={Continuous-time control synthesis under nested signal temporal logic specifications},
  author={Yu, Pian and Tan, Xiao and Dimarogonas, Dimos V},
  journal={IEEE Transactions on Robotics},
  year={2024},
  publisher={IEEE}
}

@inproceedings{lee2010geometric,
  title={Geometric tracking control of a quadrotor {{UAV}} on {{SE(3)}}},
  author={Lee, Taeyoung and Leok, Melvin and McClamroch, N Harris},
  booktitle={Proc. of CDC},
  pages={5420--5425},
  year={2010},
  organization={IEEE}
}

@inproceedings{donze2010robust,
  title={Robust satisfaction of temporal logic over real-valued signals},
  author={Donz{\'e}, Alexandre and Maler, Oded},
  booktitle={Proc. of FORMATS},
  year={2010},
}

@inproceedings{gamagedara2019geometric,
  title={Geometric controls of a quadrotor {{UAV}} with decoupled yaw control},
  author={Gamagedara, Kanishke and Bisheban, Mahdis and Kaufman, Evan and Lee, Taeyoung},
  booktitle={Proc. of ACC},
  pages={3285--3290},
  year={2019},
  organization={IEEE}
}

@book{horn2012matrix,
  title     = {Matrix Analysis},
  author    = {Horn, Roger A. and Johnson, Charles R.},
  edition   = {2},
  year      = {2012},
  publisher = {Cambridge University Press}
}

@misc{gurobi,
   author={},
  title = {{Gurobi Optimizer Reference Manual}},
  year = 2024,
  url = "https://www.gurobi.com"
}

@article{yuan2024signal,
  title={Signal Temporal Logic Planning With Time-Varying Robustness},
  author={Yuan, Yating and Quartz, Thanin and Liu, Jun},
  journal={IEEE Control Systems Letters},
  year={2024},
  publisher={IEEE}
}

@article{lee2012robust,
  title={Robust Adaptive Attitude Tracking on $\mathrm{SO}(3)$ with an Application to a Quadrotor {{UAV}}},
  author={Lee, Taeyoung},
  journal={IEEE Transactions on Control Systems Technology},
  volume={21},
  number={5},
  pages={1924--1930},
  year={2012},
  publisher={IEEE}
}

@article{invernizzi2017geometric,
  title={Geometric tracking control of a quadcopter tiltrotor {{UAV}}},
  author={Invernizzi, Davide and Lovera, Marco},
  journal={IFAC-PapersOnLine},
  volume={50},
  number={1},
  pages={11565--11570},
  year={2017},
  publisher={Elsevier}
}

@article{gu2025robust,
  title={Robust adaptive control for aggressive quadrotor maneuvers via $\mathrm{SO}(3)$ and backstepping techniques},
  author={Gu, Weibin and Primatesta, Stefano and Rizzo, Alessandro},
  journal={Robotics and Autonomous Systems},
  pages={104942},
  year={2025},
  publisher={Elsevier}
}

@ARTICLE{2020SciPy-NMeth,
  author  = {Virtanen, Pauli and Gommers, Ralf and Oliphant, Travis E. and
            Haberland, Matt and Reddy, Tyler and Cournapeau, David and
            Burovski, Evgeni and Peterson, Pearu and Weckesser, Warren and
            Bright, Jonathan and {van der Walt}, St{\'e}fan J. and
            Brett, Matthew and Wilson, Joshua and Millman, K. Jarrod and
            Mayorov, Nikolay and Nelson, Andrew R. J. and Jones, Eric and
            Kern, Robert and Larson, Eric and Carey, C J and
            Polat, {\.I}lhan and Feng, Yu and Moore, Eric W. and
            {VanderPlas}, Jake and Laxalde, Denis and Perktold, Josef and
            Cimrman, Robert and Henriksen, Ian and Quintero, E. A. and
            Harris, Charles R. and Archibald, Anne M. and
            Ribeiro, Ant{\^o}nio H. and Pedregosa, Fabian and
            {van Mulbregt}, Paul and {SciPy 1.0 Contributors}},
  title   = {{{SciPy} 1.0: Fundamental Algorithms for Scientific
            Computing in Python}},
  journal = {Nature Methods},
  year    = {2020},
  volume  = {17},
  pages   = {261--272},
  adsurl  = {https://rdcu.be/b08Wh},
  doi     = {10.1038/s41592-019-0686-2},
}

@book{lynch2017modern,
  title={Modern Robotics: Mechanics, Planning, and Control},
  author={Lynch, Kevin M. and Park, Frank C.},
  year={2017},
  publisher={Cambridge University Press}
}

@article{liu2025controller,
  title={Controller synthesis of collaborative signal temporal logic tasks for multi-agent systems via assume-guarantee contracts},
  author={Liu, Siyuan and Saoud, Adnane and Dimarogonas, Dimos V},
  journal={IEEE Transactions on Automatic Control},
  year={2025},
  publisher={IEEE}
}

@inproceedings{pant2021co,
  author={Pant, Yash Vardhan and Yin, He and Arcak, Murat and Seshia, Sanjit A.},
  booktitle={Proc. of ACC}, 
  title={Co-design of Control and Planning for Multi-rotor {{UAV}}s with Signal Temporal Logic Specifications}, 
  year={2021},
  volume={},
  number={},
  pages={4209-4216},
  doi={10.23919/ACC50511.2021.9483206}
}

@article{charitidou2024distributed,
  title={Distributed {{MPC}} with continuous-time {{STL}} constraint satisfaction guarantees},
  author={Charitidou, Maria and Dimarogonas, Dimos V},
  journal={IEEE Control Systems Letters},
  volume={8},
  pages={211--216},
  year={2024},
  publisher={IEEE}
}

@inproceedings{buyukkoccak2021distributed,
  title={Distributed planning of multi-agent systems with coupled temporal logic specifications},
  author={B{\"u}y{\"u}kko{\c{c}}ak, Ali Tevfik and Aksaray, Derya and Yazicioglu, Yasin},
  booktitle={AIAA Scitech 2021 Forum},
  pages={1123},
  year={2021}
}

@article{zou2019event,
  title={Event-triggered distributed predictive control for asynchronous coordination of multi-agent systems},
  author={Zou, Yuanyuan and Su, Xu and Li, Shaoyuan and Niu, Yugang and Li, Dewei},
  journal={Automatica},
  volume={99},
  pages={92--98},
  year={2019},
  publisher={Elsevier}
}

@article{zhou2022distributed,
  title={Distributed model predictive control for multi-robot systems with conflicting signal temporal logic tasks},
  author={Zhou, Xiaoyi and Zou, Yuanyuan and Li, Shaoyuan and Li, Xianwei and Fang, Hao},
  journal={IET Control Theory \& Applications},
  volume={16},
  number={5},
  pages={554--572},
  year={2022},
  publisher={Wiley Online Library}
}

@inproceedings{rao2023temporal,
  title={Temporal Waypoint Navigation of Multi-{UAV} Payload System using Barrier Functions},
  author={Rao, Nishanth and Sundaram, Suresh and Jagtap, Pushpak},
  booktitle = {Proc. of ECC},
  pages={1--6},
  year={2023},
  organization={IEEE}
}

@inproceedings{lindemann2019decentralized,
  title={Decentralized control barrier functions for coupled multi-agent systems under signal temporal logic tasks},
  author={Lindemann, Lars and Dimarogonas, Dimos V},
  booktitle = {Proc. of ECC},
  pages={89--94},
  year={2019},
  organization={IEEE}
}

@inproceedings{sharifi2021fixed,
  title={Fixed-time convergent control barrier functions for coupled multi-agent systems under {{STL}} tasks},
  author={Sharifi, Maryam and Dimarogonas, Dimos V},
  booktitle = {Proc. of ECC},
  pages={793--798},
  year={2021},
  organization={IEEE}
}

@inproceedings{chen2022funnel,
  title={Funnel-based cooperative control of leader-follower multi-agent systems under signal temporal logic specifications},
  author={Chen, Fei and Dimarogonas, Dimos V},
  booktitle = {Proc. of ECC},
  pages={906--911},
  year={2022},
  organization={IEEE}
}

@inproceedings{fainekos2006robustness,
  title={Robustness of temporal logic specifications},
  author={Fainekos, Georgios E and Pappas, George J},
  booktitle={International Workshop on Formal Approaches to Software Testing},
  pages={178--192},
  year={2006},
  organization={Springer}
}

@inproceedings{pant2017smooth,
  author={Pant, Yash Vardhan and Abbas, Houssam and Mangharam, Rahul},
  booktitle={2017 IEEE Conference on Control Technology and Applications (CCTA)}, 
  title={Smooth operator: Control using the smooth robustness of temporal logic}, 
  year={2017},
  volume={},
  number={},
  pages={1235-1240},
  doi={10.1109/CCTA.2017.8062628}}

@article{gilpin2020smooth,
  title={A smooth robustness measure of signal temporal logic for symbolic control},
  author={Gilpin, Yann and Kurtz, Vince and Lin, Hai},
  journal={IEEE Control Systems Letters},
  volume={5},
  number={1},
  pages={241--246},
  year={2020},
  publisher={IEEE}
}

@book{murray2017mathematical,
  title={A mathematical introduction to robotic manipulation},
  author={Murray, Richard M and Li, Zexiang and Sastry, S Shankar},
  year={2017},
  publisher={CRC press}
}

@book{khalil2002nonlinear,
  title={Nonlinear systems},
  author={Khalil, Hassan K and Grizzle, Jessy W},
  volume={3},
  year={2002},
  publisher={Prentice hall Upper Saddle River, NJ}
}

@article{storn1997differential,
  title={Differential evolution--a simple and efficient heuristic for global optimization over continuous spaces},
  author={Storn, Rainer and Price, Kenneth},
  journal={Journal of global optimization},
  volume={11},
  number={4},
  pages={341--359},
  year={1997},
  publisher={Springer}
}

@article{leung2023backpropagation,
  title={Backpropagation through signal temporal logic specifications: Infusing logical structure into gradient-based methods},
  author={Leung, Karen and Ar{\'e}chiga, Nikos and Pavone, Marco},
  journal={The International Journal of Robotics Research},
  volume={42},
  number={6},
  pages={356--370},
  year={2023},
  publisher={SAGE Publications Sage UK: London, England},
  doi={10.1177/02783649221082115}
}

\setcounter{section}{0}

\section{Proof of Lyapunov Bounds}
\label{apped:Lyapunov_Bounds}
\setcounter{equation}{0}
\renewcommand{\theequation}{I.\arabic{equation}}

The proof of the Lyapunov bounds is based on the following error dynamics and on Lemmas~\ref{lem:C_bound_new} and~\ref{lem:delta_f_Bound_new}, whose proofs are given in~\ref{apped:prof_lem2} and~\ref{apped:prof_lem3}, respectively.

For all $t\in \mathbb{T}$, let $G(t):=R_d^\top(t) R(t) \in \mathrm{SO}(3)$, and the error dynamics for \eqref{eq.e_p}--\eqref{eq.Psi(t)} are as follows.
\begin{align}
\dot{e}_p(t) &= e_v(t), \label{eq.dot_ep_app}\\
\dot{e}_v(t) &= -\frac{1}{m}(K_p e_p(t) + K_v e_v(t) - \Delta_f(t)), \label{eq.dot_ev_app}\\
\dot{\Psi}_K(t) &= e_{K,R}^\top(t) e_\omega(t), \label{eq.dot_psiK_app}\\
\dot{e}_{K,R}(t) &= \mathcal{C}(t) e_\omega(t), \label{eq.dot_eKR_app}\\
\dot{e}_\omega(t) &= -J^{-1}(e_{K,R}(t)+K_\omega e_\omega(t)), \label{eq.dot_ew_app}
\end{align}
where
\begin{align}
\mathcal{C}(t) &:= \frac{1}{2}\bigl(\mathrm{tr}[K_R G(t)]I_3-K_R G(t)\bigr), \\
\Delta_f(t)&=\left\|F_d(t)\right\|\left(\left(b_{3, d}^{\top}(t) b_3(t)\right) b_3(t)-b_{3, d}(t)\right).
\end{align}
The detailed derivation of the above error dynamics is given in~\ref{apped:deriv_errors}.
\begin{lemma}[{\cite[Proposition 2]{lee2012robust}}]
\label{lem:C_bound_new}
For all $t\in \mathbb{T}$,
\begin{equation}
 \|\mathcal C(t)\|
\le
\frac{1}{\sqrt2}\mathrm{tr}[K_R].   
\end{equation}
\end{lemma}

\begin{lemma}
\label{lem:delta_f_Bound_new}Suppose $K_R=\mathrm{diag}(k_{R_1},k_{R_2},k_{R_3})\in \mathbb{S}^3_{++}$ and
$\Psi_K(t)<\psi<h_1$. Then,
\begin{equation}
\bigl\|
b_{3,d}-b_{3,d}^\top b_3\, b_3
\bigr\|
\le
\sqrt{\frac{4g_2}{h_1}}\,
\|e_{K,R}\|.
\label{eq.b3_bound_app}
\end{equation}
which further implies that
\begin{equation}
\|\Delta_f\|
\le
\Bigl(
\|[K_p\;\;K_v]M_1^{-1/2}\|\sqrt{\mathcal V_1}
+m\|b_a\|
\Bigr)
\sqrt{\frac{4g_2}{h_1}}\,
\|e_{K,R}\|.
\label{eq.deltaf_bound_app}
\end{equation}
\end{lemma}
% \vspace*{4pt}
By \eqref{eq.dot_psiK_app}--\eqref{eq.dot_ew_app}, the time derivative of $\mathcal{V}_2(t)$ in~\eqref{eq.V_2(t)} is
\begin{align}
\dot{\mathcal V}_2(t)
&=
-c_2 e_{K,R}^\top(t) J^{-1}e_{K,R}(t)
-c_2 e_\omega^\top(t) K_\omega J^{-1}e_{K,R}(t) \notag\\
&\quad
-e_\omega^\top(t) (K_\omega-c_2\mathcal{C})e_\omega(t) .
\end{align}

Let $z_2^\top(t):=\begin{bmatrix} e_{K,R}^\top(t) & e_\omega^\top(t) \end{bmatrix}$. Then, by Lemma~\ref{lem:C_bound_new} and the definitions of $W_2$ and $\beta$ in~\eqref{eq.W_2} and~\eqref{eq.beta}, it follows that
\begin{equation}
\dot{\mathcal V}_2(t)
\le
-z_2^\top(t) W_2 z_2(t)
\le
-\beta \mathcal V_2(t).
\label{eq.V2_diff_app}
\end{equation}
By the comparison principle in~\cite[Section~3.4]{khalil2002nonlinear},
\begin{equation}
\sqrt{\mathcal V_2(t)}
\le
e^{-\beta t/2}\sqrt{\mathcal V_2(0)}.
\label{eq.V2_decay_app}
\end{equation}
Also, since $z_2^\top M_{2,1}z_2\le \mathcal V_2$, there is
\begin{equation}
\|e_{K,R}(t)\|
\le
\Bigl\|[I_3\;\;0]M_{2,1}^{-1/2}\Bigr\|
\sqrt{\mathcal V_2(t)}.
\label{eq.eKR_from_V2_app}
\end{equation}

Let $z_1^\top(t):=\begin{bmatrix} e_{p}^\top(t) & e_v^\top(t) \end{bmatrix}$. Given $\alpha_0$, $\alpha_1$, $\alpha_2$ and $\beta^{\prime}$ defined in~\eqref{eq.a0}--\eqref{eq.beta'}, let
$
C_1:=\begin{bmatrix}
\frac{c_1}{m}I_3 & I_3
\end{bmatrix}.
$
Using \eqref{eq.dot_ep_app}, \eqref{eq.dot_ev_app}, and
Lemma~\ref{lem:delta_f_Bound_new}, the time derivative of $\mathcal{V}_1(t)$ in~\eqref{eq.V_1(t)} is 
\begin{align}
\dot{\mathcal{V}}_1(t)
&=
-z_1^\top(t) W_1 z_1(t)-\Delta_f^\top(t) C_1 z_1(t) \notag\\
&\le
-\alpha_0\mathcal V_1(t)+\|\Delta_f(t)\|\,\|C_1 z_1(t)\|.
\label{eq.dot_V1_1}
\end{align}
From~\eqref{eq.V_1(t)}, it follows that $\|C_1 z_1(t)\|
\le
\|C_1 M_1^{-1/2}\|\sqrt{\mathcal{V}_1(t)}.
$
By \eqref{eq.deltaf_bound_app} and \eqref{eq.eKR_from_V2_app},
\begin{equation}
\|\Delta_f(t)\|\,\|C_1 z_1(t)\|
\le
\alpha_1\sqrt{\mathcal{V}_2(t)}\,\mathcal{V}_1(t)
+\alpha_2\sqrt{\mathcal{V}_2(t)\mathcal{V}_1(t)}.
\label{eq.bound_delf_c1z1}
\end{equation}

Substituting~\eqref{eq.bound_delf_c1z1} into~\eqref{eq.dot_V1_1} and using~\eqref{eq.V2_decay_app} yields
\begin{align}
\dot{\mathcal{V}}_1(t)
&\le
-\bigl(\alpha_0-\alpha_1 e^{-\beta t/2}\sqrt{\mathcal{V}_2(0)}\bigr)\mathcal{V}_1(t)
\notag\\
&\quad
+\alpha_2 e^{-\beta t/2}\sqrt{\mathcal{V}_2(0)}\sqrt{\mathcal{V}_1(t)}.
\end{align}
Applying the comparison principle in~\cite[Section~3.4]{khalil2002nonlinear} again yields
\begin{equation}
\sqrt{\mathcal V_1(t)}
\le
\mathcal{L}_1(\mathcal V_1(0),\mathcal V_2(0),t),
\label{eq.sqrtV1_bound_app}
\end{equation}
where
\begin{align}
&\mathcal{L}_1(\mathcal V_1(0),\mathcal V_2(0),t)
=
e^{-\alpha_0 t/2}
\Biggl[
e^{\alpha_1\sqrt{\mathcal V_2(0)}/\beta}\sqrt{\mathcal V_1(0)}
\notag\\
& \qquad \qquad \qquad
+\frac{\alpha_2\sqrt{\mathcal V_2(0)}}{2}
\int_0^t e^{(\alpha_0-\beta)s/2}\,ds
\Biggr].
\label{eq.L1_app}
\end{align}

\section{Proof of Initial Conditions}
\label{apped:IC_Bounds}
\setcounter{equation}{0}
\renewcommand{\theequation}{II.\arabic{equation}}
Define $
\mathcal{E}(t)
:=
\frac{1}{2}e_\omega^\top(t) J e_\omega(t)+\Psi_K(t)$.
With \eqref{eq.dot_psiK_app} and \eqref{eq.dot_ew_app}, it follows that
\begin{equation}
\dot{\mathcal E}(t)
=
e_\omega^\top(t) J\dot e_\omega(t)+\dot\Psi_K(t)
=
-e_\omega^\top(t) K_\omega e_\omega(t)
.
\end{equation}
Since $J, K_\omega \in \mathbb{S}^3_{++}$, $\dot{\mathcal E}(t) \le 0$, which implies that $\dot\Psi_K(t) \le 0$. With the initial conditions~\eqref{eq.Psi0<psi} and~\eqref{eq.ew0<=(1-a)psi}, it holds that
\begin{equation}
\mathcal{E}(0) = \frac{1}{2}e_\omega^\top(0) J e_\omega(0)+\Psi_K(0) < \psi < h_1.
\label{eq.init_cond_app}
\end{equation}
This indicates that $\Psi_K(t) \leq \Psi_K(0) <\psi<h_1$ for all $t\ge0$. Hence, this condition of Proposition~\ref{prop:Inequality_Psi_K} and Lemma~\ref{lem:delta_f_Bound_new} holds. Given $g_1 \|e_{K,R}(0)\|^2 \le \Psi_K(0)$ from Proposition~\ref{prop:Inequality_Psi_K}, and by $\mathcal V_2(t)$ in~\eqref{eq.V_2(t)}, one has
\begin{align}
\mathcal{V}_2(0) & =\frac{1}{2} e_\omega^{\top}(0) J e_\omega(0)+ \Psi_K(0)+c_2 e^{\top}_{K,R}(0)e_\omega(0) \notag \\
& \leq \left(1-\alpha_\psi\right)\psi+ \alpha_\psi \psi +c_2\left\|e_{K,R}(0)\right\|\left\|e_\omega(0)\right\| \notag\\
& \leq \psi + c_2 \sqrt{\frac{\psi}{g_1}}
 \left\|J^{-1 / 2}\right\|\left\|J^{1 / 2} e_\omega(0)\right\| \\
& \leq \left(1 + c_2 \sqrt{\frac{2(1-\alpha_\psi)\alpha_\psi}{\underline{\lambda}(J)g_1}}\right) \psi = \overline{\mathcal{V}}_2. \notag
\end{align}
Therefore, with $\mathcal{V}_1(0) \leq \overline{\mathcal{V}}_1$ in \eqref{eq.v10<=barV1} and $\mathcal{V}_2(0) \leq \overline{\mathcal{V}}_2$, one has $\mathcal{L}_1(\mathcal{V}_1(0), \mathcal{V}_2(0), t) \leq \mathcal{L}_1(\overline{\mathcal{V}}_1, \overline{\mathcal{V}}_2, t)$. From \eqref{eq.V_1(t)} and~\eqref{eq.sqrtV1_bound_app}, 
\begin{align}
\|e_p(t)\| &= \|\begin{bmatrix}
 \mathrm{I}_3 & 0_{3\times 3}   
\end{bmatrix}M^{-\frac{1}{2}}_1 M_1^{\frac{1}{2}}z_1(t)\| \notag\\
&\leq \|\begin{bmatrix}
 \mathrm{I}_3 & 0_{3\times 3}   
\end{bmatrix}M^{-\frac{1}{2}}_1\| \sqrt{\mathcal{V}_1(t)}\\
&\leq \|\begin{bmatrix}
 \mathrm{I}_3 & 0_{3\times 3}   
\end{bmatrix}M^{-\frac{1}{2}}_1\| \mathcal{L}_1(\overline{\mathcal{V}}_1, \overline{\mathcal{V}}_2, t) = \mathcal{L}_p(t). \notag
\end{align}

Likewise, the bound of $\|e_v(t)\|$ can be derived as
\begin{equation}
\|e_v(t)\| 
\leq \|[0_{3\times 3}  \;\; \mathrm{I}_3]M^{-\frac{1}{2}}_1\| \mathcal{L}_1\left(\overline{\mathcal{V}}_1, \overline{\mathcal{V}}_2, t\right) = \mathcal{L}_v(t).
\end{equation}

\renewcommand{\thesection}{Appendix \Roman{section}} 
\section{Multi-agent Collision Avoidance}
\label{apped:Multi-agents}
\setcounter{equation}{0}
\renewcommand{\theequation}{III.\arabic{equation}}

The radius (or span) of the $k$-th Bézier segment $B_k^i$ of the $i$-th agent quantifies that segment's spatial extent (compactness) and is defined as the maximum Euclidean distance between its geometric centroid, $\operatorname{cent}\left(B_k^i\right)$, and any of its control points:
\begin{align}
\operatorname{span}\left(B_k^i\right)=\max _{0 \leq l \leq n}\left\|c_{k, l}^i-\operatorname{cent}\left(B_k^i\right)\right\|,
\end{align}
where $\operatorname{cent}\left(B_k^i\right)=\frac{1}{n+1} \sum_{l=0}^n c_{k, l}^i$. From constraints \eqref{eq.z_epsilon},  it is proved in~\cite{yuan2024signal} that, for $\forall l\in \{1,2, \dots, n\}$,
\begin{align}
\min \left\{\left\|c^i_{k, l}-c^i_{k, 0}\right\|,\left\|c^i_{k, l}-c^i_{k, n}\right\|\right\} \leq \epsilon^i_k, 
\end{align}
which illustrates that every control point $c^i_{k, l}$ is close to at least one of the endpoints $c^i_{k, 0}$ or $c^i_{k, n}$. Denote
$
m_k^*=(c^i_{k, 0}+c^i_{k, n})/2
$
as the midpoint of the segment connecting the endpoints. Without loss of generality, assume that $c^i_{k,l}$ is closer to $c_{k, 0}$ than to $c_{k, n}$. Then,
\begin{equation}
\begin{aligned}
\left\|c^i_{k,l}-m_k^*\right\| & \leq\left\|c^i_{k,l}-c^i_{k, 0}\right\|+\left\|c^i_{k, 0}-m_k^*\right\| \\
% & \leq \epsilon_k+\left\|c^i_{k, 0}-\frac{c^i_{k, 0}+c^i_{k, n}}{2}\right\| \\
& = \epsilon^i_k+\frac{1}{2}\left\|c^i_{k, 0}-c^i_{k, n}\right\|.
\end{aligned}
\end{equation}
The same holds if closer to $c_{k, n}$. Therefore, 
\begin{equation}
\begin{aligned}
\|\operatorname{cent}(B^i_k) - m^*\| & = \|\frac{1}{n+1}\sum^{n}_{l=0} c^i_{k,l} - m^{*}\| \\
% &\leq \frac{1}{n+1}\sum^{n}_{l=0} \|c^i_{k,l}-m^{*}\| \\
% &\qquad \qquad \quad  \leq \frac{1}{n+1}\sum^{n}_{l=0} \left( \epsilon_k + \frac{1}{2}\left\|c^i_{k, 0}-c^i_{k, n}\right\|\right)\\
&\leq \epsilon_k + \frac{1}{2}\left\|c^i_{k, 0}-c^i_{k, n}\right\|
\end{aligned}
\end{equation}
% by the triangle inequality, 
Thus, for any control point $c^i_{k,l}$,  
\begin{align}
\left\|c^i_{k,l}-\operatorname{cent}\left(B^i_k\right)\right\| &\leq \left\|c^i_{k,l}-m_k^*\right\| 
+\left\|\operatorname{cent}\left(B^i_k\right)-m_k^*\right\| \notag \\
&\leq 2\epsilon^i_k + \|c^i_{k,0}-c^i_{k,n}\|.
\end{align}
yielding $\operatorname{span}(B^i_k) \leq 2\epsilon^i_k + \|c^i_{k,0}-c^i_{k,n}\|$. since any point at the Bézier curve $B^i_k(t)$ is within the convex hull, $\|B^i_k(t) - \operatorname{cent}(B^i_k)\| \leq \operatorname{span}(B^i_k)$.  
% Moreover, $\|c^i_{k,0}-c^i_{k,n}\| \leq \|c^i_{k,0}-c^i_{k,n}\|_1 $. 

Consider any two Bézier curves $B^i_{k}$ and $B^j_{k}$ over $t \in [t_k, t_{k+1}]$, with convex hulls $\mathcal{CH}^i_k = \{c^i_{k,l}\}^{n}_{l=0}$ and $\mathcal{CH}^j_k =\{c^j_{k,l}\}^{n}_{l=0}$. The constraint \eqref{eq.agents_safe} implies that
% \begin{equation}
\begin{align}
&\|\operatorname{cent}(B^i_{k}) - \operatorname{cent}(B^j_{k})\|_1 \geq 
2\epsilon^i_{k} + \|c^i_{k,0}-c^i_{k,n}\|_1 + 2\epsilon^j_{k} \notag \\
&\qquad \qquad \qquad  \qquad \quad + \|c^j_{k,0}-c^j_{k,n}\|_1 +\epsilon_{inter}\sqrt{d} \notag \\
& \qquad \qquad \quad \geq \operatorname{span}(B^i_{k}) +  \operatorname{span}(B^j_{k}) + (\epsilon_{inter}\sqrt{d}).
\end{align}
% \end{equation}

Then, by the triangle inequality, for any $t \in [t_k, t_{k+1}]$, 
\begin{equation}
\begin{aligned}    
\|B^i_{k}(t) - B^j_{k}(t)\|_1 &\geq \|\operatorname{cent}(B^i_{k})  - \operatorname{cent}(B^j_{k})\|_1 \\
& \quad - \operatorname{span}(B^i_{k}) -  \operatorname{span}(B^j_{k}) \\
& \geq \epsilon_{inter}\sqrt{d}.
\end{aligned}
\end{equation}
By the norm inequality $\|x\|_1 \leq \sqrt{d}\|x\|$, it follows that, 
\begin{equation}
\|B^i_{k}(t) - B^j_{k}(t)\| \geq \epsilon_{inter}.
\end{equation}

\section{Error Dynamics}
\label{apped:deriv_errors}
\setcounter{equation}{0}
\renewcommand{\theequation}{V.\arabic{equation}}
% $\Delta_f:=F_d-fRe_3 .$
Given the error functions~\eqref{eq.e_p}--\eqref{eq.Psi(t)}, their time derivatives are derived as follows.

{\scalebox{0.8}{$\bullet$}}\;\textbf{The derivative of $e_p(t)$:}
\begin{align}\dot{e}_p(t)=\dot{y}_{tr}(t)-\dot{y}_d(t)=e_v(t).
\end{align}

{\scalebox{0.8}{$\bullet$}}\;\textbf{The derivative of $e_v(t)$:} From~\eqref{eq.dv} and~\eqref{eq.Fd}, one has
\begin{align}
&m\dot e_v(t)
=m\ddot y_{tr}(t)-m\ddot y_d(t) \notag\\
&= -(K_p e_p(t)+K_v e_v(t)) - \left(F_d(t)-f(t)R(t)e_3\right).
\end{align}
Since $f(t)=F_d^{\top}(t)R(t)e_3$, $b_3(t)=R(t)e_3$, and
$b_{3,d}(t)=\frac{F_d(t)}{\|F_d(t)\|}$, it follows that
\begin{align}
&f(t)R(t)e_3-F_d(t) 
=\bigl(F_d^\top(t)b_3(t)\bigr)b_3(t)-\|F_d(t)\|b_{3,d}(t) \notag\\
&\qquad \qquad =\|F_d(t)\|\bigl(b_{3,d}^\top(t)b_3(t)\bigr)b_3(t)-\|F_d(t)\|b_{3,d}(t) \notag\\
&\qquad \qquad  \triangleq \Delta_f(t).
\end{align}
Thus, $
\dot e_v(t)
= -\frac{1}{m}\bigl(K_p e_p(t)+K_v e_v(t)-\Delta_f(t)\bigr).
$

{\scalebox{0.8}{$\bullet$}}\;\textbf{The derivative of ${\Psi}_K(t)$:} Let $G(t) = R^{\top}_d(t) R(t) \in \mathrm{SO}(3)$. Given $\hat{\omega}_d(t) \in \mathfrak{s o}(3)$, the following properties hold:
\begin{align}
\hat{\omega}^{\top}_d(t) &= -\hat{\omega}_d(t) \\
G(t)G^{\top}(t) &= 1 \\
G^{\top}(t) \hat{\omega}_d(t) G(t) &=\left(G^{\top}(t) \omega_d(t)\right)^{\wedge}.
\end{align}
Using these properties and the definition of $e_{\omega}(t)$, one has
\begin{equation}
\begin{aligned}
\dot G(t)
&=\dot R_d^\top(t)R(t)+R_d^\top(t)\dot R(t) \\
&=-\hat\omega_d(t)G(t)+G(t)\hat\omega(t) \\
&=G(t)\Bigl(\hat\omega(t)-\bigl(G^\top(t)\omega_d(t)\bigr)^\wedge\Bigr) \\
&=G(t)\bigl(\omega(t)-G^\top(t)\omega_d(t)\bigr)^\wedge \\
&=G(t)\hat e_\omega(t).
\end{aligned}
\end{equation}
Using the trace identity $\operatorname{tr}[A \hat{x}]=-x^{\top}\left(A-A^{\top}\right)^{\vee}$ and the definition of $e_{K, R}(t)$ yields 
\begin{equation}
\begin{aligned}
\dot{\Psi}(t) &=- \frac{1}{2} \operatorname{tr}(K_RG(t) \hat{e}_\omega(t)) \\
&= \frac{1}{2} e^{\top}_{\omega}(t)\left(K_RG(t) - (K_RG(t))^{\top}\right)^{\vee}\\
&= e^{\top}_\omega(t)e_{K,R}(t).
\end{aligned}
\end{equation}

{\scalebox{0.8}{$\bullet$}}\;\textbf{The derivative of $e_{K,R}(t)$:} Let $A(t) = K_RG(t)$, since $K_R \in \mathbb{S}^3_{++}$, $A^{\top}(t) = G^{\top}(t)K_R$. 
Using the identity $M\hat{x}+\hat{x}M^T=((\operatorname{tr}(M)\mathrm{I}-M)x)^{\wedge}$, it follows that
\begin{equation}
\begin{aligned}
\dot{e}_{K,R}(t) & =\frac{1}{2}\left(A(t)\hat{e}_\omega(t)+\hat{e}_\omega(t) A^{\top}(t)\right)^{\vee} \\
& =\frac{1}{2}\left(\operatorname{tr}(A(t))\mathrm{I}_3-A(t)\right) e_\omega(t) =\mathcal{C}(t) e_\omega(t).
\end{aligned}
\end{equation}

{\scalebox{0.8}{$\bullet$}}\;\textbf{The derivative of $e_{\omega}(t)$:} Let $G(t) = R^{\top}_d(t) R(t)$, $e_{\omega}(t)= \omega(t)-G^{\top}(t)\omega_d(t)$. Given the definition of $\dot{\omega}(t)$ and $\tau(t)$, then
\begin{equation}
\begin{aligned}
\dot{e}_\omega(t) & = \dot{\omega}(t) - \dot{G}^{\top}(t)\omega_d(t) - G^{\top}(t)\dot{\omega}_d(t) \\
& =\dot{\omega}(t)-\hat{e}_\omega^{\top}(t) G^{\boldsymbol{\top}}(t) \omega_d(t)-G^{\boldsymbol{\top}}(t) \dot{\omega}_d(t) \\
&=  J^{-1}(-\omega(t) \times J \omega(t)+\tau(t)) \\
& \quad +\hat{e}_\omega(t) G^{\boldsymbol{\top}}(t) \omega_d(t) -G^{\boldsymbol{\top}}(t) \dot{\omega}_d(t) \\
&=J^{-1}\left(-e_{K,R}(t)-K_\omega e_\omega(t)\right) \\
&\quad -\hat{\omega}(t) G^{\boldsymbol{\top}}(t) \omega_d(t)+\hat{e}_\omega(t) G^{\boldsymbol{\top}}(t) \omega_d(t)
\end{aligned}
\end{equation}

\section{Proof of Lemma 2}
\label{apped:prof_lem2}
\setcounter{equation}{0}
\renewcommand{\theequation}{V.\arabic{equation}}
The bound on $\|\mathcal C(t)\|$ is used in~\ref{apped:Lyapunov_Bounds}. We provide a brief proof using our notation, following Proposition 2 in~\cite{lee2012robust}. Let $A(t):=K_R G(t)$. Since $G(t)\in \mathrm{SO}(3)$, it follows that
\begin{equation}
\|\mathcal C(t)\|^2
\le
\frac14\bigl((\mathrm{tr}[A(t)])^2+\mathrm{tr}[A^\top(t) A(t)]\bigr).
\end{equation}
Using the cyclic property of the trace,
\begin{equation}
\mathrm{tr}[A^\top(t) A(t)]
=
\mathrm{tr}[G^\top(t)K_R^2G(t)]
=
\mathrm{tr}[K_R^2].
\end{equation}
Since $K_R=\mathrm{diag}(k_{R,1},k_{R,2},k_{R,3})$ and $|G_{ii}(t)|\le 1$,
\begin{equation}
|\mathrm{tr}[A(t)]|
=
|\mathrm{tr}[K_RG(t)]|
\le
\sum_{i=1}^3 k_{R,i}
=
\mathrm{tr}[K_R].
\end{equation}
Therefore,
\begin{equation}
\|\mathcal C(t)\|^2
\le
\frac{1}{4}\Bigl((\mathrm{tr}[K_R])^2+\mathrm{tr}[K_R^2]\Bigr)
\le
\frac{1}{2}(\mathrm{tr}[K_R])^2,
\end{equation}
which implies $
\|\mathcal C(t)\|\le \frac{1}{\sqrt2}\mathrm{tr}[K_R].
$

\section{Proof of Lemma 3}
\label{apped:prof_lem3}
\setcounter{equation}{0}
\renewcommand{\theequation}{VI.\arabic{equation}}

Let $G(t) := R_d^{\top}(t)R(t) \in \operatorname{SO}(3)$. Then,
\begin{align}
\left\|\left(b_{3,d}^{\top}(t)b_3(t)\right)b_3(t)-b_{3,d}(t)\right\|^2
&=1-\left(b_{3,d}(t)\cdot b_3(t)\right)^2 \notag \\
&=1-G_{3,3}^2(t),
\end{align}
where $G_{3,3}(t)$ denotes the $(3,3)$ entry of $G(t)$.

Let $x=[x_1,x_2,x_3]^{\top}\in\mathbb{R}^3$ be the rotation vector such that
$G(t)=\exp(\hat{x})$, where $\hat{x}\in\mathfrak{so}(3)$ is the skew-symmetric matrix associated with $x$, and let the rotation angle be $\|x\|=\theta\in[0,\pi]$. Define the unit rotation axis
$u=x/\|x\|=[u_1,u_2,u_3]^\top$. By Rodrigues' formula~\cite[Lemma~2.3]{murray2017mathematical},
\begin{equation}
G(t)=I+\sin\theta\,\hat{u}+(1-\cos\theta)\,\hat{u}^2.
\end{equation}
Hence, the $(3,3)$ entry of $G(t)$ is
\begin{equation}
\begin{aligned}
G_{3,3}(t)
&=\cos\theta+(1-\cos\theta)u_3^2 \\
&=\cos\theta+(1-\cos\theta)\frac{x_3^2}{\|x\|^2}.
\end{aligned}
\label{eq.G_33}
\end{equation}

Define $
T:=\frac{(x_1^2+x_2^2)(\cos\theta-1)}{\|x\|^2}$.
Then, by~\eqref{eq.G_33}, we have $G_{3,3}(t)=1+T$, and therefore
\begin{equation}
\begin{aligned}
1-G_{3,3}^2(t)
&=1-(1+T)^2 \\
&=-2T-T^2 \le -2T.
\end{aligned}
\label{eq.1-G_33}
\end{equation}
Since $(x_1^2+x_2^2)/\|x\|^2 \le 1$ and $1-\cos\theta\ge 0$, it follows that
\begin{equation}
\begin{aligned}
1-G_{3,3}^2(t)
&\le -2T \\
&=2\frac{(x_1^2+x_2^2)(1-\cos\theta)}{\|x\|^2} \\
&\le 2(1-\cos\theta).
\end{aligned}
\label{eq.T}
\end{equation}

From~\cite{lee2012robust}, $\Psi_K(t)$ can be expressed as
\begin{equation}
\Psi_K(t)
=\frac{1-\cos(\|x\|)}{2\|x\|^2}
\sum_{(i,j,k)\in\mathcal{C}}(k_{R,i}+k_{R,j})x_k^2,
\label{eq.Psi_K}
\end{equation}
where $\mathcal{C}=\{(1,2,3),(2,3,1),(3,1,2)\}$.

Since $
h_1=\min\{k_{R,1}+k_{R,2},\,k_{R,2}+k_{R,3},\,k_{R,1}+k_{R,3}\}$, we have
\begin{equation}
\sum_{(i,j,k)\in\mathcal{C}}(k_{R,i}+k_{R,j})x_k^2
\ge h_1\sum_{k=1}^3 x_k^2
= h_1\|x\|^2.
\end{equation}
Substituting this into~\eqref{eq.Psi_K} gives
\begin{equation}
\begin{aligned}
\Psi_K(t)
&\ge \frac{1-\cos(\|x\|)}{2\|x\|^2}\bigl(h_1\|x\|^2\bigr) \\
&=\frac{h_1}{2}\bigl(1-\cos(\|x\|)\bigr).
\end{aligned}
\end{equation}
Thus, $
1-\cos(\|x\|)\le \frac{2}{h_1}\Psi_K(t), 
$ together with~\eqref{eq.T} yields
\begin{equation}
\begin{aligned}
&\left\|\left(b_{3,d}^{\top}(t)b_3(t)\right)b_3(t)-b_{3,d}(t)\right\|^2
=1-G_{3,3}^2(t) \\
&\qquad \qquad \qquad \le 2\bigl(1-\cos(\|x\|)\bigr) \le \frac{4}{h_1}\Psi_K(t).
\end{aligned}
\end{equation}
Using the quadratic bound $\Psi_K(t)\le g_2\|e_{K,R}(t)\|^2$, one has
\begin{equation}
\left\|\left(b_{3,d}^{\top}(t)b_3(t)\right)b_3(t)-b_{3,d}(t)\right\|
\le \sqrt{\frac{4g_2}{h_1}}\,\|e_{K,R}(t)\|.
\end{equation}

Combining this with the definition of $F_d(t)$ in~\eqref{eq.Fd} and Assumption~\ref{Assump:acc_d}, the triangle inequality gives
\begin{equation}
\begin{aligned}
\|F_d(t)\|
&\le \|-K_p e_p(t)-K_v e_v(t)+mg e_3+m\ddot{y}_d(t)\| \\
&\le \|K_p e_p(t)+K_v e_v(t)\| + m\|b_a\|.
\end{aligned}
\end{equation}
Moreover, by the definition of $\mathcal{V}_1(t)$ in~\eqref{eq.V_1(t)},
\begin{align}
\|K_p e_p(t)+K_v e_v(t)\|
&=\|[K_p~K_v]z_1(t)\| \\
&\le \|[K_p~K_v]M_1^{-1/2}\|\sqrt{\mathcal{V}_1(t)}. \notag
\end{align}
Therefore, $\|\Delta_f(t)\|$ is bounded by
\begin{equation}
\begin{aligned}
&\|\Delta_f(t)\|
\le \|F_d(t)\|
\left\|\left(b_{3,d}^{\top}(t)b_3(t)\right)b_3(t)-b_{3,d}(t)\right\| \\
&\le \bigl(\|[K_p~K_v]M_1^{-1/2}\|\sqrt{\mathcal{V}_1(t)} + m\|b_a\|\bigr)
\sqrt{\frac{4g_2}{h_1}}\,\|e_{K,R}(t)\|.
\end{aligned}
\end{equation}

%                                   % the textheight of the last page by a suitable amount.
%                                   % This command does not take effect until the next page
%                                   % so it should come on the page before the last. Make
%                                   % sure that you do not shorten the textheight too much.

%%%%%%%%%%%%%%%%%%%%%%%%%%%%%%%%%%%%%%%%%%%%%%%%%%%%%%%%%%%%%%%%%%%%%%%%%%%%%%%%

%%%%%%%%%%%%%%%%%%%%%%%%%%%%%%%%%%%%%%%%%%%%%%%%%%%%%%%%%%%%%%%%%%%%%%%%%%%%%%%%



% \section*{ACKNOWLEDGMENT}

% The preferred spelling of the word ÒacknowledgmentÓ in America is without an ÒeÓ after the ÒgÓ. Avoid the stilted expression, ÒOne of us (R. B. G.) thanks...Ó  Instead, try ÒR. B. G. thanksÓ. Put sponsor acknowledgments in the unnumbered footnote on the first page.

% %%%%%%%%%%%%%%%%%%%%%%%%%%%%%%%%%%%%%%%%%%%%%%%%%%%%%%%%%%%%%%%%%%%%%%%%%%%%%%%%

% References are important to the reader; therefore, each citation must be complete and correct. If at all possible, references should be commonly available publications.

\end{document}